\documentclass[aps,prb,twocolumn,showpacs,floatfix,eqsecnum]{revtex4}

\pdfoutput=1

\usepackage{amsmath}
\usepackage{graphicx}
\usepackage{epsfig}
\usepackage{wrapfig}

\newcommand{\bsub}{\begin{subequations}}
\newcommand{\esub}{\end{subequations}}

\newcommand \bea {\begin{eqnarray} }
\newcommand \eea {\end{eqnarray}}
 
\newcommand{\beg}{\begin{equation}}
\newcommand{\en}{\end{equation}}
\newcommand{\bp}{\mathbf p}
\newcommand{\bq}{\mathbf q}

\newcommand \bel  {\begin{align}}
\newcommand \enl  {\end{align}}

\newcommand{\vecr}{\vec r}
\newcommand{\bk}{\mathbf k}
\newcommand{\up}{\uparrow}
\newcommand{\dn}{\downarrow}
\newcommand{\dg}{^\dagger}

\begin{document}

\title{Non-universal weak antilocalization effect in cubic topological Kondo
insulators}

\author{Maxim Dzero$^{1,2}$, Maxim G. Vavilov$^3$, Kostyantyn Kechedzhi$^{4}$ and Victor M. Galitski$^{5,6}$}
\affiliation{$^1$Department of Physics, Kent State University, Kent, OH 44242, USA}
\affiliation{$^2$Max Planck Institute for Physics of Complex Systems, N\"{o}thnitzer str. 38, 01187 Dresden, Germany}
\affiliation{$^3$Department of Physics, University of Wisconsin, Madison, Wisconsin 53706, USA}
\affiliation{$^4$QuAIL and USRA, NASA Ames Research Center, Mail Stop 269-3, Moffett Field, CA 94035, USA}
\affiliation{$^5$Joint Quantum Institute and Condensed Matter Theory Center, Department of Physics, University of Maryland, College Park, Maryland 20742, USA}
\affiliation{$^6$School of Physics, Monash University, Melbourne, Victoria 3800, Australia}

\begin{abstract} 
We study the quantum correction to conductivity on the surface of cubic topological Kondo insulators with multiple Dirac bands. We consider the model of time-reversal invariant disorder which induces the scattering of the electrons within the Dirac bands as well as between the bands. When only intraband scattering is present we find three long-range diffusion modes which lead to weak antilocalization correction to conductivity, which remains independent of the microscopic details such as Fermi velocities and relaxation times. Interband scattering gaps out two diffusion modes leaving only one long-range mode. We find that depending on the value of the phase coherence time, either three or only one long-range diffusion modes contribute to weak localization correction rendering the quantum correction to conductivity non-universal. We provide an interpretation for the results of the recent transport experiments on samarium hexaboride where weak antilocalization has been observed. 
\end{abstract}

\date{\today}

\pacs{72.15.Qm, 73.23.-b, 73.63.Kv, 75.20.Hr}

\maketitle

%\tableofcontents

\section{Introduction}
Samarium hexaboride (SmB$_6$) along with PuB$_6$ and YbB$_6$ have recently 
emerged as prominent candidates\cite{Dzero2010,Takimoto2011,Dzero2012,Deng2013,XiDai2013,XiDai2014} for hosting topologically protected metallic surface states.\cite{HasanKane2010} In particular, SmB$_6$ - a material in which strong hybridization between samarium conduction $d$-electrons and strongly correlated $f$-electrons drives an onset of an insulating state
at low temperatures\cite{Allen1979,Cooley1995,Nyhus1997,Riseborough2000,Caldwell2007} - has recently came into focus of 
theoretical and experimental studies as a most prominent candidate for the first correlated topological insulator.\cite{Wolgast2013,Kim2013,Zhang2013,GLi2013,Hatnean2013,Kim2014,Baruselli2014,Phelan2014,Xu2014,RadiationDamage1,NernstSmB6}

In order to experimentally establish the existence of the helical conduction time-reversal invariant states on a surface of a generic topological insulator using either transport or thermodynamic measurements, one needs to show that (i) the conduction at low temperatures is limited to the surface; (ii) an inclusion of a small amount of magnetic impurities leads to localization of the surface states, i.e. presence of the time-reversal breaking scattering potentials leads to localization; (iii) single particle states have 
linear momentum dispersion along the surface and no dispersion in the normal to the surface direction and (iv) 
there is a strong spin-orbit interaction, which leads to coupling between the momentum and spin of the conduction electron giving rise to the helicity of the carriers. In samarium hexaboride, a series of state-of-the-art transport studies have unambiguously shown that the resistivity plateau at temperatures below 5 Kelvin\cite{Allen1979,Cooley1995} is governed by surface conduction only.\cite{Wolgast2013,Kim2013,Kim2014} Furthermore, Kim, Xia and Fisk \cite{Kim2014} have examined the low-temperature
transport properties of the alloys Sm$_{1-x}A_x$B$_6$ for the non-magnetic yttrium and ytterbium ($A$=Y,Yb) and magnetic gadolinium ($A$=Gd) substitutions. They found that while small amount ($\sim 3\%$) of gadolinium leads to insulating behavior in resistivity, substitutions of non-magnetic ions do not cause destruction of the metallic surface states.\cite{Kim2014} To verify that the conduction electrons on the surface have Dirac dispersion, G. Li {et al.} \cite{GLi2013} have experimentally studied the quantum oscillations of magnetization under applied external magnetic field. By plotting the dependence of the index $n$ which labelled the positions of the maxima in magnetization versus the inverse of magnetic field, G. Li et al. have shown that there is a contribution corresponding to the zero energy state which would only be possible for conducting state with Dirac-like dispersion. 

\begin{figure}[htb]
\includegraphics[width=7.15cm,height=7.15cm]{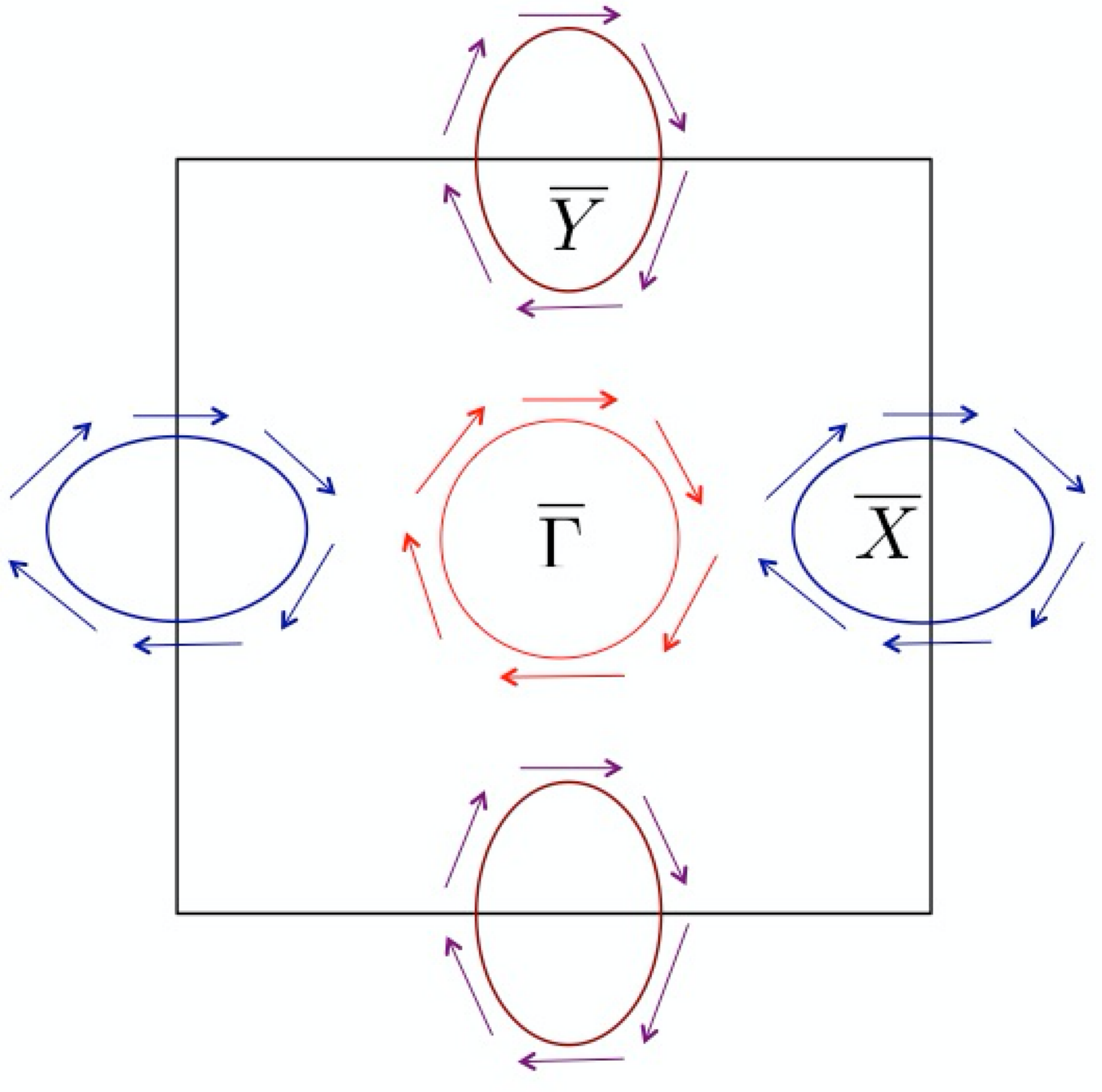}
\caption{(Color online) Surface band structure for the strong topological insulator emerging from the inversion of the odd and even parity bands at the $X$ points of the bulk Brillouin zone. Arrows denote the spin texture of the surface carriers corresponding to the ground state configuration with the $\Gamma_8$ quartet of the $f$-levels.\cite{Baruselli2014,Xu2014} We assume that chemical potential crosses all three bands. Without loss of generality we consider the bands with the same chirality. In addition we will neglect the ellipticity of the Dirac pockets, but take
into account the difference in the Fermi velocities of the electrons in different pockets.}
\label{Fig2Bands}
\end{figure}
%mv: Maxim, can you make the vertical axis to have marks starting with 0.5 or even with the range 0 to 2.
% MD: done
\begin{figure}[htb]
\includegraphics[width=8.25cm,height=6.45cm]{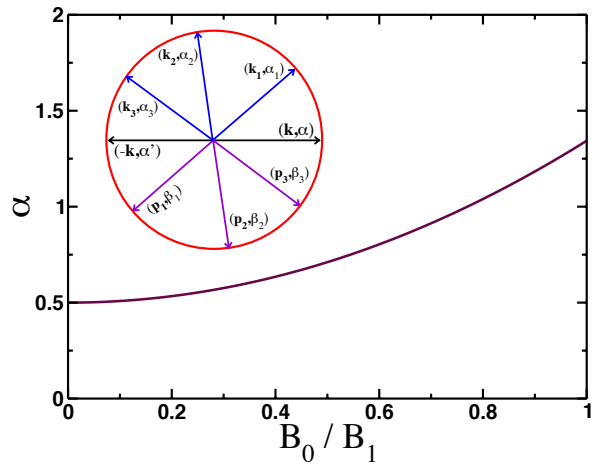}
\caption{(Color online) Schematic plot for the crossover behavior of the dimensionless coefficient $\alpha$, appearing in Hikami-Larkin-Nagaoka expression (\ref{HLN}), as a function of the ratio $B_0/B_1$ Eq. (\ref{alphaexp}) for the quantum correction to conductivity for the surface states in topological Kondo insulators. The value of the remaining parameter is chosen as $B_2=0.9B_1$. Within the three-band model for the topological surface states, in the absence of the interband scattering there are three diffusion modes, which govern the quantum correction to conductivity with $\alpha\sim 3/2$. However, in the presence of the interband scattering processes two out of three diffusion modes become gapped, so that $\alpha\sim 1/2$. However, at moderately high temperatures when $\tau_\phi$ becomes comparable to the gap of the remaining two diffusion modes also contribute leading to $\alpha\sim 3/2$. Inset shows two independent scattering processes  from state with momentum $\bk$ in band $\alpha$ to a state with momentum $-\bk$ in band $\alpha'$ leading to the quantum correction to conductivity.}
\label{Fig1Crossover}
\end{figure}

The magnitude of the spin-orbit interaction for the surface electrons in topological insulators can be indirectly probed by studying the quantum interference correction to conductivity:\cite{Hikami1980,HZLu2011,Evelina2011,Lu2011PRB,Garate2012} 
%mv: the paragraph contains a repetition, can be shortened.
upon decrease in temperature, $\delta T<0$, increase in conductivity 
%mv: ($\delta\sigma>0$)
($\delta\sigma>0$) would signal weak anti-localization effect as opposed to weak localization corresponding to decrease in conductivity 
($\delta\sigma < 0$). The sign of the correction to conductivity is determined by the ratio of the spin-orbit scattering length, $l_{SO}$, to the dephasing length $l_{\phi} $:
%mv: \gg l$ (here $l$ is a mean-free path): 
%for $\delta T<0$ and 
for weak spin-orbit coupling, $l_{SO}\gg l_\phi$ (here $l$ is a mean-free path) correction to conductivity is negative, 
%mv: correction to conductivity is negative, 
while in the opposite limit of
strong spin-orbit coupling $l_{SO}\ll l_\phi$ and the interference correction to conductivity is positive.
 
Thus, helicity of the Dirac-like carriers on the surface of topological insulators, associated with the fixed polarization of electron spin perpendicular to the momentum direction (i.e. $l_{SO}\sim l$, here $l$ is a mean-free path), necessarily leads to weak-antilocalization due additional Berry phase acquired by the carriers as they scatter along the time-reversed paths.\cite{QiZhang2011,HZLu2011} 
%mv: add refs for original Lso papers, similar to Hikami et al but with T dependence, or the nexr paragraph is Ok?  

Application of an external magnetic field perpendicular to conducting surface destroys quantum interference processes leading to positive or negative magneto-conductivity -- 
%mv: added 
another
signature of weak localization or weak anti-localization. The corresponding magnetic field dependence of the conductivity correction is then described by famous Hikami-Larkin-Nagaoka (HLN) formula:\cite{Hikami1980}
\beg\label{HLN}
\Delta\sigma_{\textrm{HLN}}(B)=\frac{\alpha e^2}{2\pi^2\hbar}\left[\log\frac{B_0}{B}-\psi\left(\frac{1}{2}+\frac{B_0}{B}\right)\right],
\en
where $\alpha$ is a dimensionless parameter determined by the number of conduction channels and the strength of the spin-orbit coupling, $\psi$ is the digamma function, $B_0=\hbar/4el_\phi^2$. Moreover, $\alpha>0$ for the case of the strong spin-orbit coupling and each independent conduction channel contributes $1/2$ to the value of $\alpha$, so that $\alpha=1$ for the case of Rashba-split bands on the surface. For the topological surface states, $\alpha=1/2$ for the case of a single Dirac band and $\alpha=3/2$ for the three Dirac bands. 

Recently, S. Thomas et al. \cite{Thomas2013WAL} have studied weak anti-localization effect and magneto-conductivity in samarium hexaboride.  By fitting the experimental data with HLN formula, Eq. (\ref{HLN}) for the case of a single band, the value of the parameter $\alpha$ came out to be approximately equal to one, $\alpha\approx 1$, 
%mv:  I thought they were consistent with \alpha=1/2, i.e. topological surface states?
for several sets of data. 
In the most recent transport experiments by Y. Nakajima et al. \cite{JP1d} also observe  weak-antilocalization effect and their results seem to be generally in agreement with the earlier studies. 
%mv: On one hand t
On the other hand, these observations would contradict a natural expectation that the value of alpha must be close to $\alpha\approx 3/2$, which is due to the three Dirac bands on the surface of SmB$_6$.
\cite{Takimoto2011,XiDai2013,Alexandrov2013,Roy2014}. 
%mv: On the other hand, 

%mv: Lastly, i
%
%mv: Motivated by these experimental results, i

%mv: combined text and removed: conduction channels which would contribute to interference correction to conductivity thus reducing the value to $\alpha\sim1/2$. 
Motivated by these experimental results, we calculate the quantum interference correction to conductivity in a generic cubic topological Kondo insulator. Our main goal is to account for the fairly wide distribution of values for the parameter $\alpha$ obtained by 
analyzing the experimental data. \cite{Thomas2013WAL,JP1d}.
As we argue in this paper, the presence of the disorder-induced interband scattering suppresses two antilocalization modes and reduces the value of $\alpha$ from naive $\alpha=$ [1/2 $\times$ number of zones].
We consider the surface band structure which consists of three Fermi pockets: one at surface $\overline{\Gamma}$ point and two at $\overline{X}$ and $\overline{Y}$ points
of the two-dimensional Brillouin zone, Fig. \ref{Fig2Bands}. We assume that the electron scattering within each band - intraband scattering - provides the strongest scattering mechanism, so that elastic scattering time is the shortest time scale in the problem. Consequently, the disorder scattering between various bands is considered as a correction to the intraband one. We show that (i) for the case when only intraband disorder time-reversal invariant scattering is present, all three conduction channels (per surface) will contribute to the interference correction to conductivity; (ii) the inclusion of the interband scattering shows that two conduction channels are suppressed while the remaining one contributes to weak anti-localization effect. However, at higher temperatures when the inverse dephasing time 
$\tau_\phi^{-1}\sim T^{p}$ ($p>0$) becomes comparable with the size of the gap in the spectrum of the diffusion modes, we still find that all three modes contribute to conductivity. By considering $\tau_\phi$ as a parameter, we describe this crossover behavior by showing the dependence of the parameter $\alpha$ on $\tau_\phi$ schematically on Fig. \ref{Fig1Crossover}. For the analysis of magnetoconductivity 
using HLN formula (\ref{HLN}) our result implies that the size of the correction for small fields and the large fields generally would correspond to different values of $\alpha$. Specifically, we will show that the (\ref{HLN}) generalizes to 
\beg\label{HLN3}
\Delta\sigma(B)=\frac{e^2}{2\pi^2\hbar}\sum\limits_{i=0}^2\alpha_i\left[\log\frac{B_i}{B}-\psi\left(\frac{1}{2}+\frac{B_i}{B}\right)\right],
\en
where $\alpha_i$ are the dimensionless parameters, which depend on the diffusion coefficients of the surface electrons from each band and
scattering times and $B_{1,2}$ are determined by the interband scattering times, which give rise to the gap in the long-range diffusion modes, 
%mv:
while $B_{i=0} =  \hbar/4el_\phi^2$ is still defined by the inelastic dephasing length $l_\phi$. 
In the limit of infinite interband scattering times - $B_{1,2}$ become equal to $B_0$ and we recover the HLN formula. In the limit of low magnetic field $B_i/B\gg 1$ (\ref{HLN3}) simplifies to 
\beg\label{HLN3SmallB}
\Delta\sigma(B\to 0)\approx-\frac{e^2}{24\pi^2\hbar}\sum\limits_{i=0}^2{\alpha_i}\left(\frac{B}{B_i}\right)^2.
\en
If we now compare this expression with (\ref{HLN}) in the limit of small magnetic fields, we see that parameter $\alpha$ can be written
as
\beg\label{alphaexp}
\alpha=\sum\limits_{i=0}^2{\alpha_i}\left(\frac{B_0}{B_i}\right)^2
\en
We see that in the case when temperatures are not very low, $B_0\sim B_i$ and 
%mv: even the gapped 
even the gapped diffusion modes will nearly equally contribute to the localization correction.
Therefore, depending on the temperature at which the experiments are performed and on the surface disorder one,
expects that the values of the parameter $\alpha$ extracted by fitting the experimental data using (\ref{HLN}) may vary from $\alpha\sim 1/2$ to $\alpha\sim 3/2$ for the case of three Dirac bands. This situation is schematically shown on Fig. \ref{Fig1Crossover}.

Our paper is organized as follows. In Section II we introduce the model for the surface states. 
In Section III we present the calculation of the interference correction to conductivity for the uncorrelated mixture of the 
intraband and interband disorder potentials. In Section IV we generalize the results from the previous Sections to the case
of the correlated mixture of the scattering potentials. Sections V and VI are devoted to 
the discussion of our results and conclusions. Throughout the paper we adopt the energy units $\hbar=c=1$.

\section{Surface states in the presence of disorder}

In this Section we setup the model and introduce the parameters which will be used in the calculation for the interference correction to conductivity. 
\subsection{Hamiltonian and correlation functions}
The Hamiltonian for the surface electrons in cubic topological Kondo insulators can be written as a sum of three terms, which describe electrons near $\overline{\Gamma}$, $\overline{X}$ and $\overline{Y}$ points in the 2D Brillouin zone, Fig. \ref{Fig2Bands}:
\cite{Takimoto2011,XiDai2013,Alexandrov2013,Roy2014}
\beg\label{Eq1}
\hat{H}=\sum\limits_{j=\Gamma, X,Y} \sum\limits_{\bp\sigma}\psi_{j\bp\sigma}\dg v_j({\vec \sigma}\cdot{\vec p})\psi_{j\bp\sigma}.
\en
Here we neglect the anisotropies in velocities along $x$ and $y$-direction in $\overline{X}$ and $\overline{Y}$ pockets.\cite{Takimoto2011,Roy2014} In Eq. (\ref{Eq1}) momentum is taken relative to the center of the pocket. Introducing the six component spinor
\beg
\hat{\Psi}^T=[\psi_{\Gamma\bp\up} ~\psi_{\Gamma\bp\dn} ~\psi_{X\bp\up} ~\psi_{X\bp\dn} ~\psi_{Y\bp\up} ~\psi_{Y\bp\dn}]
\en
the Hamiltonian can be compactly written as follows
\beg\label{H}
\hat{H}=v{\vec \Sigma}\cdot{\vec p}, \quad {\vec \Sigma}=\Pi_v\otimes{\vec \sigma}, \quad 
\Pi_v=\left(\begin{matrix} {\zeta_\Gamma} & 0 & 0 \\ 0 & {\zeta_X} & 0 \\ 0 & 0 & {\zeta_Y} \end{matrix}\right)
\en
and the coefficients $\zeta_{\Gamma,X,Y}=v_{\Gamma,X,Y}/v$ account for velocity anisotropies on different pockets. The underlying cubic symmetry requires $v_X=v_Y$, however there is no symmetry constrains on the ratio of velocities at $\Gamma$ and $X,Y$ pockets, so that generally $v_\Gamma\not=v_{X,Y}$.
In addition, we define the retarded and advanced Green's functions:
\beg\label{GAR}
\begin{split}
&\check{G}_0^{R,A}(\bp,\epsilon)=\left[
\begin{matrix} \hat{G}_{0\Gamma}^{R,A}(\bp,\epsilon) & & 0 \\  
 & \hat{G}_{0X}^{R,A}(\bp,\epsilon) & & \\
0 & & \hat{G}_{0Y}^{R,A}(\bp,\epsilon)
\end{matrix}
\right], \\ 
&\hat{G}_{0j}^{R,A}(\bp,\epsilon)=\frac{(\epsilon\pm i\delta)\hat{\sigma}_0+v_j({\vec \sigma}\cdot{\vec p})}{(\epsilon\pm i\delta)^2-v_j^2p^2}.
\end{split}
\en
Note that in the Hamiltonian (\ref{H}) we ignore the higher order terms in momentum as well as other type of conduction
channels which may arise due to polarity driven bands,\cite{ZHZhu2013} strong surface potential \cite{Roy2014} etc. We will discuss how the presence of  additional terms in the Hamiltonian may affect our results in Section \ref{Discussion}.

\subsection{Intraband disorder}

In the following we construct a theory of the metallic conductance of the disordered surface state of SmB$_6$ using perturbative expansion in $p_Fl\gg1$. The values estimated from experiments $p_Fl\sim 100$ suggest the perturbation theory to be a good approximation for the data~\cite{NernstSmB6}.   
We consider $N_i$ impurities with potential $u_0$ and matrix structure described by $\hat{U}$ leading to the 
following expression 
\beg\label{Vdis}
\begin{split}
&\hat{V}({\mathbf r})=u_0\hat{U}\sum\limits_{i}\delta({\mathbf r}-{\mathbf R}_i).
\end{split}
\en
Averaging over an impurity ensemble yields
\beg\label{average}
\begin{split}
&\hat{V}(\bp)=
\sum\limits_iu_0\hat{U}e^{i\bp\cdot{\mathbf R}_i}, \quad \langle{\hat{V}(\bp)}\rangle_{dis}=0, \\
&\langle{\hat{V}(\bp_1)\hat{V}(\bp_2)}\rangle_{dis}=n_i{\cal A}u_0^2\hat{U}\hat{U}\delta(\bp_1+\bp_2),
\end{split}
\en
where ${\cal A}$ is the surface area and $n_i=N_i/{\cal A}$. In the following we first consider the intraband disorder,
which implies that matrix $\hat{U}$ has the following block-diagonal structure:
\beg\label{Udia}
\hat{U}=\hat{I}_{3\times3}\otimes\hat{\sigma}_0,
\en
where $\hat{I}_{3\times 3}$ is a unit matrix. We first evaluate the disorder averaged correction to the self-energy:
\beg\label{Sigma}
\check{\Sigma}(\epsilon)=n_iu_0^2\int\frac{d^2\bp}{(2\pi)^2}\hat{U}\check{G}_0^{R,A}(\bp,\epsilon)\hat{U}
\en
Clearly, $\check{\Sigma}(\epsilon)$ has a block-diagonal matrix structure similar to the expression for the retarded and advanced propagators (\ref{GAR}).
It follows
\beg\label{taua0}
\begin{split}
\hat{\Sigma}_j(\epsilon)&\approx\mp
\frac{i\hat{\sigma}_0}{2\tau_{j0}}, \quad \tau_{j0}^{-1}=\pi n_i\nu_ju_0^2.
\end{split}
\en
where $\tau_{j0}$ is elastic scattering rate, we neglected the real part since it leads to a small correction to $\epsilon\approx v_jp_{Fj}$, $p_{Fj}$ are the corresponding Fermi momenta and
$\nu_j$ is a single particle density of states per spin $\nu_j=p_{Fj}/2\pi v_j$.
The corresponding expressions for the renormalized retarded and advanced correlators become
\beg\label{GRGA}
\hat{G}_{j}^{R,A}(\bp,\epsilon)=\frac{(\epsilon\pm i/2\tau_{j0})\hat{\sigma}_0+v_j({\vec \sigma}\cdot{\vec p})}{(\epsilon\pm i/2\tau_{j0})^2-v_j^2p^2}.
\en
Similar expression for the correlation functions has been found for graphene. \cite{McCann2006}
\subsection{Classical conductivity}
Next, we can use the expressions for the correlators (\ref{GRGA}) to calculate the classical conductivity, which is
given by  
\beg\label{sigma_ab}
\sigma_{\alpha\beta}(\omega)=\frac{e^2}{2\pi}\int\frac{d^2\bp}{(2\pi)^2}\textrm{Tr}\left\{\hat{v}_\alpha\check{G}^R(\bp,\epsilon+\omega)
\hat{v}_\beta\check{G}^A(\bp,\epsilon)\right\}.
\en
Here $\hat{v}_\alpha$ are the components of the velocity defined by the momentum derivative of the Hamiltonian (\ref{H})
$\hat{v}_\alpha={\vec \nabla}_{p_\alpha} \hat{H}=v\hat{\Sigma}_\alpha.$
The block-diagonal structure of matrices entering into (\ref{sigma_ab}) allows one to write
\beg\label{sigma_ab2}
\begin{split}
\sigma_{\alpha\beta}^{(0)}(\omega)&\approx\delta_{\alpha\beta}\sum\limits_j\frac{e^2\nu_jv_j^2\tau_{j0}}{2(1-i\omega\tau_{j0})}
\end{split}
\en
where the subscript denotes that we have neglected the vertex corrections and on the last step we have assumed $\epsilon\gg \{\omega,\tau_{j0}^{-1}\}$.  The vertex corrections
can be formally included by making the following substitution in (\ref{sigma_ab}):
\beg
\hat{v}_\alpha\to\Lambda\int\frac{d^2\bp}{(2\pi)^2}\hat{U}\check{G}^A(\epsilon,\bp)\hat{v}_\alpha\check{G}^R(\epsilon+\omega,\bp)\hat{U}
\en
Summing the resulting geometric series to all orders we obtain
\beg\label{sigma0ab}
\sigma_{\alpha\beta}(\omega)=\sum\limits_j\frac{e^2\nu_jv_j^2\tau_{\textrm{tr},j}}{2(1-i\omega\tau_{\textrm{tr},j})}, \quad \tau_{\textrm{tr},j}=2\tau_{j0}.
\en
Thus, we find for the Dirac electrons the transport lifetime for the conducting states is twice the elastic scattering time.\cite{McCann2006}

\begin{figure}[htb]
\includegraphics[width=8.25cm,height=2.15cm]{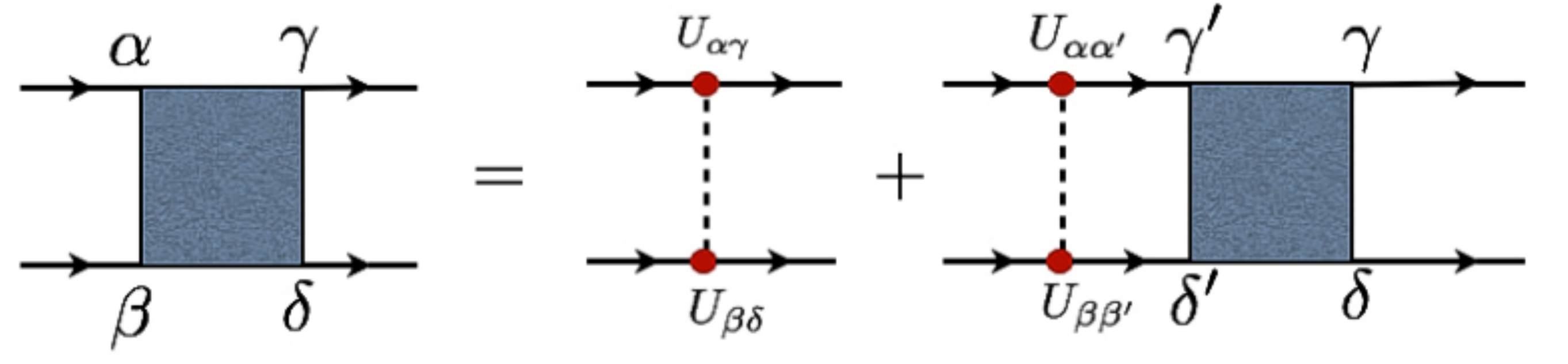}
\caption{(Color online) Dyson equation for the Cooperon propagator $\Gamma_{\alpha\beta,\gamma\delta}(\omega,\bq)$. Here the 
Greek indices encode both spin and Dirac cone - "valley" - components, solid lines represent the single particle propagators and $\hat{U}$ accounts for the disorder potential.}\label{FigCooperon}
\end{figure}

\subsection{Cooperon Propagator}

The quantum interference correction to conductivity (\ref{sigma0ab}) 
is associated with the two-particle correlation function known as
the Cooperon. It satisfies the Bethe-Salpeter equation which is represented diagrammatically on Fig. \ref{FigCooperon}.
In what follows, we will use separate notations for the "valley" and "spin" indices: we adopt latin superscripts for valley $a=1,2,3$ and Greek subscripts for components of the Kramers doublet. Equation for the Cooperon propagator reads:
\beg\label{Gamma2}
\begin{split}
&\Gamma_{\alpha\beta,\gamma\delta}^{ab,cd}(\omega,\bq)=\Lambda_0 U_{\alpha\gamma}^{ac}U_{\beta\delta}^{bd}\\&+
\Lambda_0\sum\int\frac{d^2\bp}{4\pi^2}U_{\alpha\alpha'}^{aa'}[\check{G}^R]_{\alpha'\gamma'}^{a'c'}(\epsilon+\omega,\bp)\\&\times\Gamma_{\gamma'\delta',\gamma\delta}^{c'd',cd}(\omega,\bq)U_{\beta\beta'}^{bb'}
[\check{G}^A]_{\beta'\delta'}^{b'd'}(\epsilon,\bq-\bp),
\end{split}
\en
with $\Lambda_0=n_iu_0^2$ and the summation goes over repeated indices. Since $\check{G}_{\alpha\beta}^{A,R}$ are diagonal in "valley" indices, we have
\beg
[\check{G}^{A,R}]_{\alpha\beta}^{ab}(\epsilon,\bp)=[\hat{G}_a^{R,A}(\epsilon,\bp)]_{\alpha\beta}\delta_{ab}
\en
Next for convenience we introduce the spin singlet and triplet components of the Cooperon\cite{Rainer1985,McCann2006} and define
\beg\label{Cs1s2ToGamma}
\begin{split}
C_{S_1S_2}^{ab,cd}(\omega,\bq)&=\frac{1}{2}\sum(\sigma_y\sigma_{S_1})_{\alpha\beta}
\Gamma_{\alpha\beta,\gamma\delta}^{ab,cd}(\omega,\bq)(\sigma_{S_2}\sigma_{y})_{\delta\gamma},
\end{split}
\en
where $S_{1,2}=0,x,y,z$. Using the following identities 
\beg\label{Firtz}
\begin{split}
&\sum\limits_{S=0,x,y,z}(\sigma_{S}\sigma_{y})_{\alpha\beta}(\sigma_y\sigma_{S})_{\mu\nu}=2\delta_{\alpha\nu}\delta_{\beta\mu}, \\
&\sum\limits_{a=x,y,z}[\sigma_{a}]_{\alpha\beta}[\sigma_a]_{\mu\nu}=2\delta_{\alpha\nu}\delta_{\beta\mu}-\delta_{\alpha\beta}\delta_{\mu\nu},
\end{split}
\en
we can now express the components of Cooperon (\ref{Gamma2}) in the right hand side in terms of matrices (\ref{Cs1s2ToGamma}) using
\beg\label{GammaToCs1s2}
\Gamma_{\alpha\beta,\gamma\delta}^{ab,cd}(\omega,\bq)=\frac{1}{2}\sum\limits_{S_1,S_2}(\sigma_{S_1}\sigma_{y})_{\beta\alpha}
C_{S_1S_2}^{ab,cd}(\omega,\bq)(\sigma_y\sigma_{S_2})_{\gamma\delta},
\en
which follows directly from relations (\ref{Cs1s2ToGamma}) and (\ref{Firtz}).
To obtain the equation for the singlet and triplet components of the Cooperon, we multiply both parts of the equation (\ref{Gamma2}) by the product of Pauli matrices introduced in (\ref{Cs1s2ToGamma}).

\section{Intraband Scattering}
%\subsection{intraband scattering}

\subsection{Gapless cooperon modes}
\begin{figure}[htb]
%mv: what if we put these figure side by side? will notations be too small?
\includegraphics[width=7.25cm,height=10.15cm]{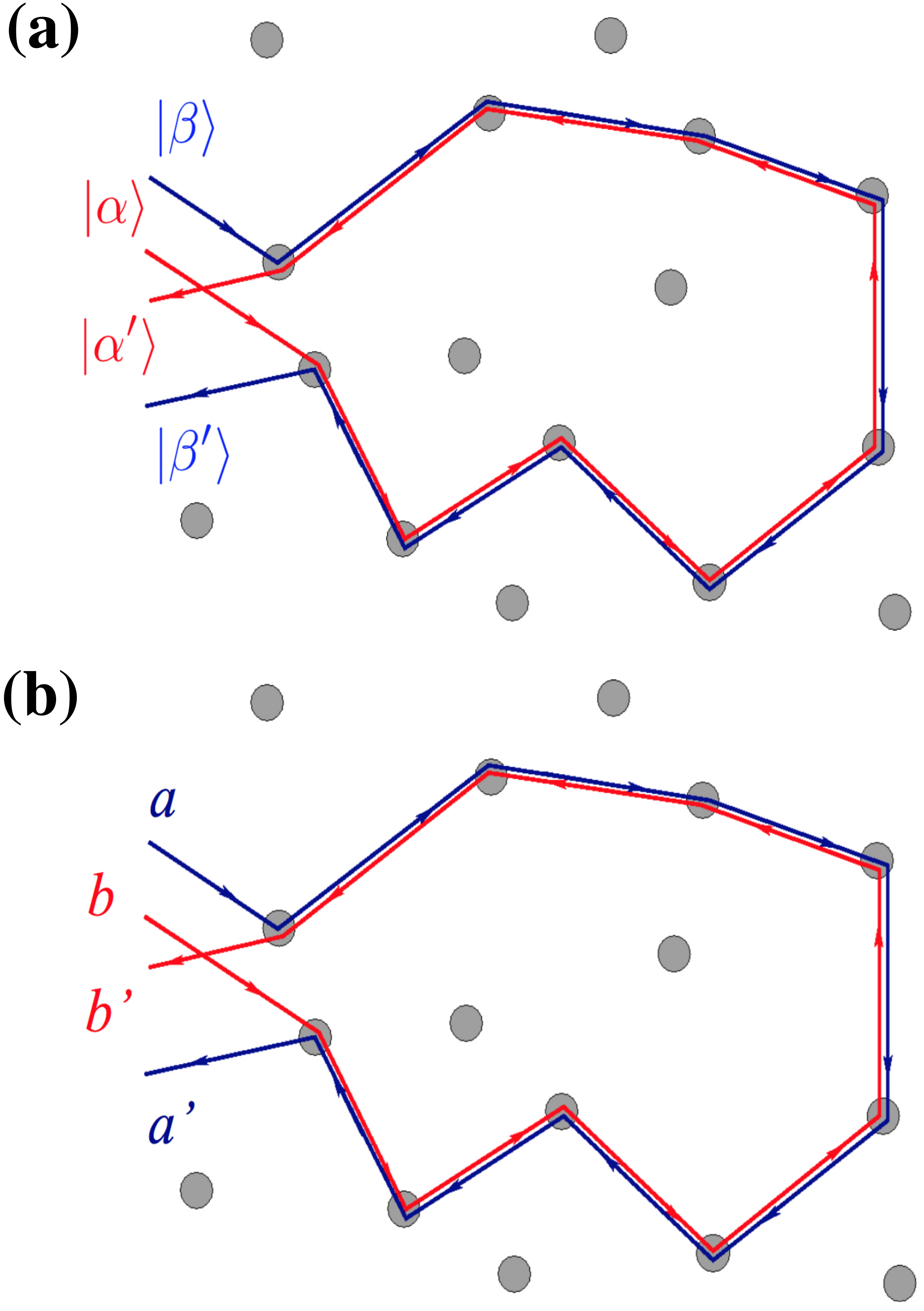}
\caption{(Color online) (a) Pseudospin is coupled to momentum via strong spin-orbit coupling. As a result only one mode
$(|\up\dn\rangle-|\dn\up\rangle)/\sqrt{2}$ corresponding to a total pseudospin $S=0$ state - singlet mode -  survives;
(b) Propagation of electrons from different bands is completely uncorrelated as bands have different Fermi energies and velocities (see text for details). As a result we only need to consider $a=b$ case. Even in the case of the interband scattering, the propagation of the chiral electrons still remains coherent.}
\label{Fig3Scatter}
\end{figure}

%mv
To keep our calculations transparent, in this section we analyze the system with intraband scattering only and postpone treatment of the interband scattering for the next section.
Here the matrix for the intraband disorder potential is diagonal in both spin and valley indices
(\ref{Udia}). In the momentum integral we insert equation (\ref{GammaToCs1s2}). 
This gives an overall prefactor of $(1/2)^2$ in front of the second term, but the subsequent trace over the product of Pauli matrices will cancel one of them. It follows
\beg\label{Cs1s2}
\begin{split}
&C_{S_1S_2}^{ab,a'b'}(\omega,\bq)={\Lambda}_0\delta_{aa'}\delta_{bb'}\delta_{S_1S_2}\\+&\frac{\Lambda_0}{2}\sum
\int\frac{d^2\bp}{4\pi^2}(\sigma_y\sigma_{S_1})_{\alpha\beta}[\hat{G}_a^R(\epsilon+\omega,\bp)]_{\alpha\gamma'}\\&\times
[\hat{G}_b^A(\epsilon,\bq-\bp)]_{\beta\delta'}
(\sigma_{S}\sigma_{y})_{\delta'\gamma'}C_{SS_2}^{ab,a'b'}(\omega,\bq)
\end{split}
\en
Clearly, there are two possibilities: (i) when the retarded and advanced propagators belong to the same valley, i.e. $a=b=a'=b'$ and (ii) when they belong to different bands or valleys, $a\not=a', b\not=b'$. 
To evaluate the trace under the integral we recall the definition of the Greens functions (\ref{GRGA}) which are diagonal in the band index $a,b$. Electric conductivity is given by a trace over band indexes in Eq.~(\ref{sigma_ab}) and therefore in the absence of 
interband scattering only Cooperon components that are diagonal in band indexes %mv: Cooperons 
contribute to quantum corrections to conductivity. For a diagonal Cooperon each term in Eq.~(\ref{Cs1s2}) is diagonal in band indexes and therefore the system of equations splits into a set of independent equations for each band.
%Despite the fact that there is no disorder potential, which could have scattered electrons between the different valleys, the Cooperon matrix does not retain the block diagonal form in valley %ndices. However, due to the mismatch between impurity scattering times and velocities, the off-diagonal in valley indices components of the inverse Cooperon matrix are
%of the order of $O(1/\epsilon\tau_{a0})$. We make the following conjecture: the off-diagonal components of the inverse Cooperon matrix will not affect the singular nature of the Cooperon modes arising from the scattering within the same valleys. 
 Thus, we only need to consider the components $C_{S_1S_2}^{a}(\omega,\bq)$ defined in the same band, Fig. \ref{Fig3Scatter}. After short calculation we find that the equation for the components of the Cooperon matrix in the valley $a=\Gamma,X,Y$ can be compactly written as follows:
\beg\label{Eq4C}
\hat{M}_a\cdot\hat{C}^{a}=\Lambda_a\hat{I}_{4\times 4}.
%\left(\begin{matrix} 1 & 0 & 0 & 0 \\ 0 & 1 & 0 & 0 \\ 0 & 0 & 1 & 0 \\ 0 & 0 & 0 & 1 \end{matrix}\right), 
\quad \Lambda_a=\tau_{a0}^{-1}\Lambda_0
\en
and the elements of the matrix $\hat{M}$ are given by Eq. (\ref{M}) in Appendix \ref{MIntra}. In limit of $\omega\tau_{a0}\ll 1$ and $v_aq\ll 1$ for the matrix $\hat{M}_a$ we find:
\beg\label{M}
\hat{M}_a=\left[
\begin{matrix}
\frac{v_{a}^2q^2\tau_{a0}}{2}-i\omega & -\frac{i}{2}v_aq_x &-\frac{i}{2}v_aq_y & 0 \\
-\frac{i}{2}v_aq_x & \frac{1}{2}\left(\frac{1}{\tau_{a0}}-i\omega\right) & 0 & 0 \\
-\frac{i}{2}v_aq_y & 0 & \frac{1}{2}\left(\frac{1}{\tau_{a0}}-i\omega\right) &  0 \\
 0 & 0 & 0 &\frac{1}{\tau_{a0}}
\end{matrix}
\right].
\en
To find the components of the Cooperon matrix we need to find an inverse of the matrix $\hat{M}_a$. The quantum correction
to conductivity is determined by the diagonal components $C_{ii}^a$ of the Cooperon matrix. The singlet component is given by 
\beg\label{C00}
\begin{split}
C_{00}^a(\omega,q)&=\frac{\Lambda_0}{v_a^2\tau_{a0}^2q^2-i\omega\tau_{a0}},
\end{split}
\en
while for the three triplet components we get
\beg\label{Ctriplet}
\begin{split}
C_{xx}^a(\omega,q)&=\Lambda_0\frac{v_a^2q^2\tau_{a0}^2(1+\sin^2\phi_q)-2i\omega\tau_{a0}}{v_a^2\tau_{a0}^2q^2-i\omega\tau_{a0}},\\
C_{yy}^a(\omega,q)&=\Lambda_0\frac{v_a^2q^2\tau_{a0}^2(1+\cos^2\phi_q)-2i\omega\tau_{a0}}{v_a^2\tau_{a0}^2q^2-i\omega\tau_{a0}}, \\
C_{zz}^a(\omega,q)&=\Lambda_0.
\end{split}
\en
Thus we find that for the \emph{intraband disorder} out of four Cooperon modes per each Fermi pocket, only one mode - singlet modes $C_{00}^a(\omega,q)$ - remains gapless: propagation of electrons from Fermi pockets remains uncorrelated. 
This is not surprising given the strong spin-orbit coupling for the surface electrons. 

\subsection{WAL correction to conductivity}\label{WALIntraBand}
As we have just demonstrated, the singlet channel is the most singular one (\ref{C00ab}), so that we can ignore the correction to conductivity arising from the triplet components. Thus, for the singular weak localization correction (see Appendix B for details) we find
\beg\label{TotalCorrection_nointer}
\begin{split}
\delta\sigma(0) \approx\sum\limits_{a=\Gamma,X,Y}&{
e^2\nu_av_a^2\tau_{a0}^3}
\int C_{00}^a(\omega,\bq)\frac{d\bq}{(2\pi)^2}.
\end{split}
\en
Carrying out the momentum integral, we find
\beg\label{CorrectionFin_nointer}
\begin{split}
\delta\sigma(0)&=\frac{e^2}{4\pi^2}
\sum\limits_{a=\Gamma,X,Y}
\log\left(\frac{\tau_\phi}{\tau_{a0}}\right),
\end{split}
\en
where $\tau_\phi$ is the phase coherence time. Since $\delta\sigma(0)>0$
the inteference correction leads to weak antilocalization. Lastly, we remind the reader that on symmetry grounds $\tau_{X0}=\tau_{Y0}\not=\tau_{\Gamma0}$. 

In an external magnetic field, the momentum integral should be replaced with the sum over Landau levels. Specifically, in the presence of the perpendicular magnetic field $B$, the momentum is quantized: $q_n^2=(n+1/2)l_B^{-2}$ with $l_B^2=1/4eB$.
Setting $-i\omega=\tau_\phi^{-1}$ in the expression for the Cooperon and performing the summation over Landau levels yields the following expression for the magneto-conductivity $\delta\sigma(B)$:
\beg\label{LHNSmB6_nointer}
\begin{split}
&\delta\sigma(B)= \frac{e^2}{2\pi^2}
\\&\times
\sum\limits_{a=\Gamma,X,Y} \frac{1}{2}\left\{\Psi\left[\frac{1}{2}+\frac{B_0}{B}\left(\frac{l_{\phi a}}{l_a}\right)^2\right]-\Psi\left[\frac{1}{2}+\frac{B_0}{B}\right]\right\},
\end{split}
\en
where we took into account the leading (gapless) contribution to the Cooperon, $l_{\phi a}^2=D_a \tau_\phi=1/4eB_0$, $l_a^2=D_a\tau_{a0}$ and $\Psi(z)$ is the digamma function, and we assume that $\tau_\phi$ is the same for all bands. The first digamma function in this expression can be replaced with $\log(B_0l_{\phi a}^2/Bl_a^2)$ due to the large value of its argument and the $\log(l_{\phi a}^2/l_a^2)$ 
%mv:
drops out of 
$\Delta\sigma(B)=\delta\sigma(B)-\delta\sigma(0)$. As a result, we can immediately identify the pre-factor $\alpha$ in Eq.~(\ref{HLN}) from Eq.~\eqref{LHNSmB6_nointer}  as $\alpha=3/2$.
This value of $\alpha$ is a simple consequence of the  fact that for a three band model without interband scattering 
%mv: where 
each band gives a universal contribution to the WAL correction to the conductivity.

\section{Interband disorder}

Now we consider the disorder potential which also includes the component which induces 
the scattering between the different pockets: we expect the latter to remove the divergent nature of some of the singlet Cooperon components $C_{00}$. Without loss of generality we will consider perhaps the simplest type of the interband disorder potential:
\beg\label{Uinter}
\begin{split}
&V(\vecr)=u_0\sum\limits_i\hat{U}_0\delta(\vecr-\vecr_i)+u_{x}\sum\limits_{j}\hat{U}_{x}\delta(\vecr-\vecr_j),
\\ &\hat{U}_0=\hat{\tau}_0\times\hat{\sigma}_0, \quad \hat{U}_{x}=\hat{T}_x\times\hat{\sigma}_0.
\end{split}
\en
Here, $\hat{T}_x$ is a $3\times 3$ matrix whose diagonal elements all equal to zero, while off-diagonal elements equal to one:
$[\hat{T}_x]_{ab}=(1-\delta_{ab})$. Just as in the case of the weak antilocalization in graphene,\cite{McCann2006} one can show that this
disorder potential captures the essential physical features needed to describe the long-range diffusion modes. 
Moreover, we assume that any given impurity does not scatter electrons simultaneously between different pockets and within the same pocket - no correlations between the impurity scattering - so that for disorder average we have
\beg\label{disav}
\begin{split}
\langle V(\vecr_1)V(\vecr_2)\rangle_{dis.}&=\Lambda_0\hat{U}_0\hat{U}_0\delta(\vecr_1-\vecr_2)\\&+\Lambda_{x}\hat{U}_{x}\hat{U}_{x}\delta(\vecr_1-\vecr_2),
\end{split}
\en
where $\Lambda_0=n_{i0}u_0^2$ and $\Lambda_{x}=n_{ix}u_{x}^2$ with $n_{ix}$ being the concentration of the interband scatterers. We define the corresponding scattering times
\beg\label{sctime}
\begin{split}
\tau_{ab}^{-1}=\pi n_{ix}u_{x}^2\nu_{b},~\tau_{a}^{-1}=\tau_{a0}^{-1}+\sum\limits_{b\not=a}\tau_{ab}^{-1},
\end{split}
\en
with the second expression following directly from Eqs. (\ref{Sigma},\ref{Uinter},\ref{disav}), 
while intraband scattering time $\tau_{a0}$ is defined by Eq. (\ref{taua0}). The notations we adopt for the scattering times in (\ref{sctime}) should be understood as follows: for $a,b=\Gamma,X,Y$ %$a=\Gamma$, $b=X,Y$. 
The remaining scattering times are obtained by cycling permutation of these indices. Note, that our choice of the interband disoroder potential leads to 
$\tau_{\Gamma X}=\tau_{YX}$ but at the same time $\tau_{\Gamma X}\not=\tau_{X\Gamma}$. The corresponding self-energy correction to the 
Green's function now reads
\beg\label{GARNew}
\hat{\Sigma}_a^{R,A}(\epsilon)\approx\mp \frac{i\hat{\sigma}_0}{2\tau_{a}}.
\en

\subsection{Cooperon modes in presence of interband scattering}
% with uncorrelated disorder}

As we have already mentioned above our goal in this section is to verify if the gapless singlet Cooperon modes acquire a gap due to 
interband scattering effects.  In what follows we will ignore the triplet Cooperon components - they are gapped already - and obtain the equation for the singlet Cooperon components only. The corresponding equation for the pseudospin components (in the valley space) of the Cooperon matrix can be obtained similarly to Eq. (\ref{Cs1s2}). It follows: 
\beg\label{IntraCooperon}
\begin{split}
&C_{00}^{ab,a'b'}(\omega,\bq)=
{\Lambda}_0\delta_{aa'}\delta_{bb'}+{\Lambda}_x[\hat{U}_x]_{aa'}[\hat{U}_x]_{bb'}+\\&+\sum\limits_{a''b''}{\cal M}_{aa'',bb''}(\omega,\bq)C_{00}^{a''b'',a'b'}(\omega,\bq),
\end{split}
\en
where we introduced the following matrix  
\beg\label{Ma1a2b1b2}
\begin{split}
&{\cal M}_{aa',bb'}(\omega,\bq)=
\frac{\Lambda_0}{2}\sum
\int\frac{d^2\bp}{4\pi^2}(\sigma_y)_{\alpha\beta}\\&\times
[\hat{G}_a^R(\epsilon+\omega,\bp)]_{\alpha\gamma'}[\hat{G}_b^A(\epsilon,\bq-\bp)]_{\beta\delta'}
(\sigma_{y})_{\delta'\gamma'}\delta_{aa'}\delta_{bb'}+\\&+\frac{\Lambda_x}{2}
\sum
\int\frac{d^2\bp}{4\pi^2}[\hat{U}_x]_{aa'}(\sigma_y)_{\alpha\beta}[\hat{G}_{a'}^R(\epsilon+\omega,\bp)]_{\alpha\gamma'}\\&\times
[\hat{G}_{b'}^A(\epsilon,\bq-\bp)]_{\beta\delta'}
(\sigma_{y})_{\delta'\gamma'}[\hat{U}_x]_{bb'}
\end{split}
\en
for convenience and summations are performed over repeated spin and pseudospin indices.

%Since there are three bands in the problem the singlet Cooperon matrix is generally $12\times 12$ and last term in Eq. (\ref{IntraCooperon}) equation accounts for the intraband scattering. Its contribution is largest for $a'=b'$, since the terms with $a'\not =b'$ are of the order $1/p_Fl\ll 1$ smaller.

In presence of inter-band scattering it is important to take into account the off-diagonal elements of the Cooperons in the band space, i.e. consider the full $9\times9$ matrix (with spin indexes $S=S_1=0$).
This consideration is significantly simplified due to the effect of the Fermi line missmatch between different bands, due to $\Gamma$ and $X,Y$ bands having different Fermi velocities. Furthermore, $X$ and $Y$ bands are characterized by asymmetric Fermi lines, see Fig.~\ref{Fig2Bands}. The resulting missmatch of the phases of wave functions results in: (i) suppression of the disorder induced interband scattering matrix element due to the Fermi wavelength missmatch; (ii) suppression of the contribution of the interband terms to the conductivity $\propto 1/((p_a-p_b) l)$, where $p_a, p_b$ are Fermi momenta for bands $a$ and $b$; (iii) suppression of the interband interference Cooperon modes with $a\neq b$ or $c\neq d$ in Eq.~(\ref{IntraCooperon}) due to Fermi line assymetry. 
%mv: in fig 4, we use a,b and a', b', in equation 3.15 abcd. Can we make c->a''; d->b''? or just single  prime but use a different notation for dummy indices?  
The latter effect is somewhat analogous to the Fermi line trigonal warping effect in the band structure of graphene~\cite{McCann2006} and is present even in the case of very small Fermi line missmatch with an important distinction that in SmB$_6$ considered here it suppresses interband interference. In SmB$_6$ it is likely that $(p_a-p_b) l\gg 1$ for all bands and therefore the mechanism (ii) is significant. Formally this means that in the case of interband terms in the Cooperon and Hikami boxes calculated in Appendix~\ref{MIntra}~and~\ref{ConductivityCorrections} would be smaller by factor of $  1/((p_a-p_b) l$ as compared to the ones corresponding to intraband terms. Therefore, we only need to consider nine elements in the Cooperon matrix: three diagonal ones $C_{00}^{aa,aa}\equiv C_{00}^a$ and six off-diagonal ones: $C_{00}^{aa,bb}\equiv C_{00}^{ab}$. Consequently, we introduce the following notations:
\beg\label{C00ab}
\hat{C}\approx
\left(
\begin{matrix}
C_{00}^{\Gamma} & C_{00}^{\Gamma X} & C_{00}^{\Gamma Y} \\
C_{00}^{X\Gamma} & C_{00}^{X} & C_{00}^{X Y} \\
C_{00}^{Y\Gamma} & C_{00}^{YX} & C_{00}^{Y} 
\end{matrix}
\right)
\en 

\paragraph{Eigenvalues of the Cooperon matrix.} To solve the equation (\ref{IntraCooperon}) we first consider the following eigenvalue problem:\cite{Pikusetal1998}
\beg\label{Eigen1}
\begin{split}
\lambda_i\Psi_{ab}^{(i)}=\sum\limits_{a'b'}{\cal M}_{aa',bb'}(0,0)\Psi_{a'b'}^{(i)},
\end{split}
\en
where the components of the matrix $\hat{\cal M}$ are given by (\ref{Ma1a2b1b2}). The quick calculation (see Appendix A for details) shows that 
${\cal M}_{aa',bb'}(0,0)={\cal M}_{aa'}\delta_{ab}\delta_{a'b'}$ with
\beg\label{Maap}
\hat{\cal M}=\left(
\begin{matrix}
\frac{\tau_\Gamma}{\tau_{\Gamma0}} & \frac{\tau_X}{\tau_{\Gamma X}} & \frac{\tau_Y}{\tau_{\Gamma Y}} \\
\frac{\tau_\Gamma}{\tau_{X\Gamma}} & \frac{\tau_X}{\tau_{X0}} & \frac{\tau_Y}{\tau_{XY}} \\
\frac{\tau_\Gamma}{\tau_{Y\Gamma}} & \frac{\tau_X}{\tau_{YX}} & \frac{\tau_Y}{\tau_{Y0}} \\
\end{matrix}
\right).
\en
The eigenvalues (\ref{Eigen1}) of this matrix can be most compactly written in terms of the following parameters
\beg\label{Params4Eigen}
\begin{split}
r_{ab}&=\frac{1}{2}\left(\frac{\tau_a}{\tau_{ab}}\frac{\tau_b}{\tau_{ba}}-t_at_b\right), ~t_{a}=\frac{1}{2}\left(\frac{\tau_{a}}{\tau_{a0}}-1\right)\\
\gamma_\pm&=-\sum\limits_at_a\pm\sqrt{\left(\sum\limits_at_a\right)^2+\sum\limits_{a\not=b}r_{ab}}
\end{split}
\en
with $t_{a}=\frac{\tau_{a}}{2\tau_{a0}}-\frac{1}{2}$. Note that since the parameters $r_{ab}<0$ and $t_a<0$ 
the parameters $\gamma_{\pm}$ are positive.
Then the expressions for the eigenvalues (\ref{Eigen1}) are
\beg\label{Res4Eigen}
\lambda_0=1, \quad \lambda_{1,2}=1-\gamma_{\pm}, 
\en
Clearly, if we neglect the interband scattering, we find a threefold degenerate eigenvalue $\lambda=1$. As we have discussed above this situation corresponds to the existence of three gapless modes. Inclusion of the interband scattering processes lifts the degeneracy leaving only one gapless mode.  Below, we first show that $\gamma_{\pm}$ determine the gap for the diffusion modes and then 
compute the diffusion coefficient for the gapless mode. 
\paragraph{Eigenvectors.} From the analysis of the equation (\ref{Eq4Aij}) it is clear that the diverging contribution to the Cooperon emerges for the eigenvalue $\lambda_0=1$ with the corresponding components of an eigenvector
\beg\label{Psi0}
\Psi_{ab}^{(0)}=N_{0}\frac{\delta_{ab}}{\tau_a}, 
\en
where $a=\Gamma,X,Y$ and proportionality coefficient $N_{0}$ is the normalization constant. Similarly, the eigenvector components for an eigenvalue $\lambda=1-\gamma_{+}$ are
\beg\label{Psi12}
\Psi_{ab}^{(1)}=\frac{N_{1}}{(1+\gamma_a)\tau_a}\delta_{ab}-
\left(\sum\limits_{d}\frac{N_{0}N_{1}}{(1+\gamma_d)\tau_d^2}\right)\Psi_{ab}^{(0)}.
\en
Here $N_{1}$ is a normalization constant and $\gamma_a=\gamma_{+}\tau_a^{-1}/(\tau_{\Gamma X}^{-1}+\tau_{\Gamma Y}^{-1}+\tau_{X\Gamma}^{-1})$. Similar expression can be obtained for the third eigenvector $\Psi_{ab}^{(2)}$.  
We note, that the eigenvectors $\Psi_{ab}^{(i)}$ satisfy the orthonormalization condition, 
\beg\label{NormA}
\sum\limits_{a,b}{\Psi}_{ab}^{(i)}\Psi_{ab}^{(j)}=\delta_{ij}.
\en

\begin{figure}[htb]
\includegraphics[width=8.25cm,height=6.25cm]{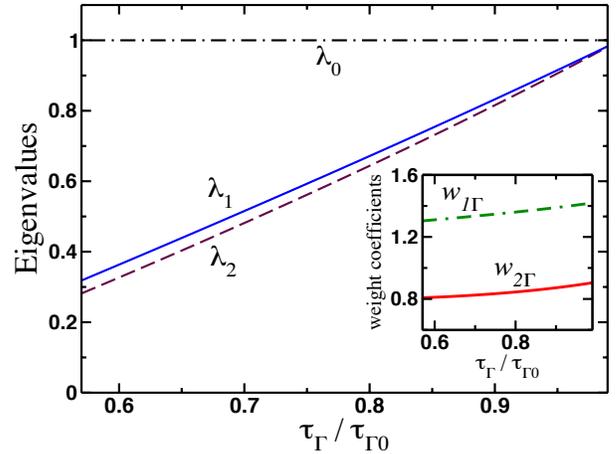}
\caption{(Color online)  Plot of the eigenvalues $\lambda_i$ ($i=0,1,2$) and the weight coefficients which appear in Eq. (\ref{CooperonA0j})
as a function of $\tau_\Gamma/\tau_{\Gamma 0}$. Here without loss of generality we consider a special case when the following relation 
between the scattering times holds: $\frac{\tau_a}{\tau_{ba}}=\frac{\tau_{b}}{\tau_{ab}}$ ($a\not=b$)}
\label{FigEigen}
\end{figure}

Having computed the eigenvectors, we can now express the components of the Cooperon (\ref{Eigen1}):
\beg\label{Coop2Psi}
C_{00}^{ab,cd}(\omega,\bq)=\sum\limits_{ij}{\cal A}_{ij}(\omega,\bq)\Psi_{ab}^{(i)}{\Psi}_{cd}^{(j)},
\en
Using this equation together with the normalization condition (\ref{NormA}), we can re-write the equation for the components of the Cooperon matrix (\ref{IntraCooperon}) as follows:
\beg\label{Eq4Aij}
\sum\limits_{k}\left[(1-\lambda_i)\delta_{ik}+{\cal W}_{ik}(\omega,\bq)\right]{\cal A}_{kj}(\omega,\bq)={\cal V}_{ij},
\en
where the matrix elements ${\cal W}_{ik}(\omega,\bq)$ are given by 
\beg\label{Wik}
\begin{split}
&{\cal W}_{ik}(\omega,\bq)=\\&=\sum\limits_{ab,a'b'}[{\cal M}_{aa',bb'}(0,0)-{\cal M}_{aa',bb'}(\omega,\bq)]\Psi_{a'b'}^{(i)}{\Psi}_{ab}^{(k)}
\end{split}
\en
and they are obviously vanishing for $\bq\to0$ and $\omega\to0$. The matrix elements ${\cal V}_{ij}$ are obtained from expanding the first two terms in
the right hand side of the Eq. (\ref{IntraCooperon}) in terms of the eigenvectors $\Psi_{ab}^{(i)}$ :
\beg\label{Vij}
{\Lambda}_0\delta_{ac}\delta_{bd}+{\Lambda}_x[\hat{U}_x]_{ac}[\hat{U}_x]_{bd}=\sum\limits_{ij}{\cal V}_{ij}\Psi_{ab}^{(i)}{\Psi}_{cd}^{(j)}.
\en
It is straightforward to compute the coefficients ${\cal A}_{ij}(\omega,\bq)$ by solving the system of linear equations (\ref{Eq4Aij}). 
For example, the coefficients ${\cal A}_{0j}(\omega,\bq)$, which contribute to the gapless Cooperon mode can be found by using the fact that the matrix elements ${\cal W}_{ik}(\omega,\bq)$ are small for small $\omega$ and $\bq$. It follows
\beg\label{A0j}
{\cal A}_{0j}(\omega,\bq)\approx\frac{{\cal V}_{0j}}{{\cal W}_{00}(\omega,\bq)}.
\en
The matrix elements entering into this expression can be found from (\ref{Vij},\ref{Wik}) using the normalization condition (\ref{NormA}). However, for our subsequent analysis of the quantum correction to conductivity we will also need the remaining two contributions to the Cooperon.
After some algebra we find that the resulting expression for the diagonal components of the Cooperon (\ref{C00ab},\ref{Coop2Psi}) can be written as follows:
\beg\label{CooperonA0j}
C_{00}^{a}(\omega,\bq)=\frac{\tau_t\Lambda_0}{\tau_{a}^2}\sum\limits_{i=0}^{2}\frac{w_{ia}}{Dq^2-i\omega+\Gamma_{ia}},
\en
where 
\beg\label{transporttau}
\frac{1}{\tau_t}=\sum_a\frac{1}{\tau_a}
\en
and the gaps in the denominator are
\beg\label{CoopGaps}
\Gamma_{0a}=0, \quad \Gamma_{1a}=\tau_{a}^{-1}\gamma_{+}, \quad \Gamma_{2a}=\tau_{a}^{-1}\gamma_{-}
\en
and $w_i\sim O(1)$ are dimensionless parameters determined by the combination of the intra- and interband scattering rates. For reader's convenience, on Fig. \ref{FigEigen} we plot the dependence of the coefficients $w_{1\Gamma}$ and $w_{2\Gamma}$ and the eigenvalues $\lambda_i$ as the function of the ratio $\tau_\Gamma/\tau_{\Gamma 0}$ for the specific choice of the scattering times such that $\tau_a/\tau_{ba}=\tau_b/\tau_{ab}$. 

In the expression for the singlet Cooperon we introduced the diffusion coefficient for the singlet long-range mode:
\beg\label{LamD}
\begin{split}
D=\sum\limits_{a}\left[\frac{\tau_t}{\tau_{a0}}D_a+\frac{\tau_t}{\tau_a}\sum\limits_{b\not=a}\frac{\tau_b}{\tau_{ab}}D_b\right]
\end{split}
\en
Thus, we find that pseudospin symmetry breaking (band mixing) perturbations produce the relaxation in the spin-singlet components of the Cooperon except for the single mode, which is protected by the time-reversal symmetry.\cite{McCann2006} The trajectories in Fig.~\ref{Fig3Scatter}(b) giving rise to the gapless Cooperon mode correspond to each ballistic segment the blue and red lines (moving in opposite directions shown by arrows) reside in the same electron band in the BZ and combine to form a spin-singlet. This is the pair of trajectories the interference between which is protected by time reversal, since time reversal maps each band on itself. Note that this is in contrast to the case of graphene where the time-reversal symmetry maps the two different bands on each other and therefore the band-singlet Cooperon mode is protected by time reversal symmetry~\cite{McCann2006}.
In the following Section we evaluate the quantum correction to conductivity appearing due to the presence of this single long-range diffusion mode focusing specifically on the value of the pre-factor $\alpha$ appearing in the HLN formula (\ref{HLN}).

\subsection{WAL correction to conductivity}

% duplicated text, see above, may need more editing

We again disregard the correction to conductivity arising from the triplet components as they are suppressed by the intraband scattering with the largest gap scale in our model. Moreover, we can omit gapped singlet modes due to the interband scattering keeping only the gapless mode identified in the previous section.  This gapless mode is protected from dephasing due to the disorder scattering, but is still suppressed on length scales $l_\phi=\sqrt{D\tau_\phi}$ due to the inelastic electron scattering characterized by decoherence time $\tau_\phi $.
We have, see Appendix B for details:
\beg\label{TotalCorrection}
\begin{split}
\delta\sigma\approx\sum\limits_{a=\Gamma,X,Y}&{\left(2-\frac{\tau_{a}}{\tau_{a0}}\right)e^2\nu_av_a^2\tau_a^3}\times\\&\times
\int C_{00}^a(\omega,\bq)\frac{d\bq}{(2\pi)^2}.
\end{split}
\en
%mv : did it stay from previous versions:  with $i=x,y$. 
Carrying out the momentum integral, we find
\beg\label{CorrectionFin}
\begin{split}
\delta\sigma(0)&=\frac{e^2}{2\pi^2}\left[\sum\limits_{a}\frac{D_a\tau_t}{2D\tau_{a0}}\left(2-\frac{\tau_{a}}{\tau_{a0}}\right)\right]\\
&\times\sum\limits_{i=0}^2w_{ia}\log\left(\frac{\tau_t^{-1}}{\textrm{max}\{\tau_\phi^{-1},\Gamma_{ia}\}}\right),
\end{split}
\en
where $\tau_\phi$ is the phase coherence time and we assumed $\tau_t^{-1}\gg\gamma_{\pm}\tau_a^{-1}$ ($i=1,2$). Since $\delta\sigma(0)>0$
the inteference correction leads to WAL. 

Similarly to the calculation in Section~\ref{WALIntraBand} in presence of an external magnetic field the summation over Landau levels yields the following expression for the magneto-conductivity $\delta\sigma(B)$:
\beg\label{LHNSmB6}
\begin{split}
&\delta\sigma(B)=\frac{e^2}{2\pi^2}\left[\sum\limits_{a}\frac{w_{0a}D_a\tau_t}{2D\tau_{a0}}\left(2-\frac{\tau_{a}}{\tau_{a0}}\right)\right]\\&\times\left\{\Psi\left[\frac{1}{2}+\frac{B_0}{B}\left(\frac{l_\phi}{l_t}\right)^2\right]-\Psi\left[\frac{1}{2}+\frac{B_0}{B}\right]\right\},
\end{split}
\en
where we took into account the leading (gapless) contribution to the Cooperon, $l_\phi^2=D\tau_\phi=1/4eB_0$, $l_t^2=D\tau_t$ and $\Psi(z)$ is the digamma function. The first digamma function in this expression can be replaced with $\log(B_0l_\phi^2/Bl_t^2)$ 
%mv:
for  $l_\phi/l_t\gg 1$ and we recover the structure of the HLN expression~(\ref{HLN}) for $\Delta\sigma(B)=\delta\sigma(B)-\delta\sigma(0)$.  
Therefore, from our result for the correction to conductivity (\ref{LHNSmB6}) we can immediately identify the pre-factor in front of the square bracket with the the weight factor $\alpha$ in Eq. (\ref{HLN}):
\beg\label{alpha}
\alpha=\sum\limits_{a}\frac{w_{0a}D_a}{2D}\left(2-\frac{\tau_{a}}{\tau_{a0}}\right)\frac{\tau_t}{\tau_{a0}}.
\en

For the moderately large magnetic fields, however, the remaining two modes will contribute to conductivity. Their contribution is formally given by 
the same expression as (\ref{LHNSmB6}) where we have to replace $\tau_\phi^{-1}$ with $\gamma_\pm\tau_a^{-1}$. It follows
\beg\label{LHNSmB62}
\begin{split}
\Delta\sigma_{\textrm{gap}}(B)&\approx\frac{e^2}{2\pi^2}\sum\limits_{i=1}^2\frac{\alpha w_{ia}}{w_{0a}}\\&\times\left[\log\frac{B_i}{B}-\psi\left(\frac{1}{2}+\frac{B_i}{B}\right)\right]
\end{split}
\en 
with $4eB_i=(\gamma_\pm/D)\cdot\textrm{max}\{\tau_\Gamma^{-1},\tau_X^{-1},\tau_Y^{-1}\}$, so that the total correction to conductivity becomes $\Delta\sigma_{\textrm{tot}}=\Delta\sigma+\Delta\sigma_{\textrm{gap}}$. Note, that unlike (\ref{LHNSmB6}), the contribution from 
$\Delta\sigma_{\textrm{gap}}$ is temperature independent. 

As we have already discussed above, in the absence of the interband scattering, the eigenvalue problem (\ref{Eigen1}) becomes degenerate, since in that case the second term on the right hand side of that equation vanishes, while the integral in the first term equals one. Therefore, in that case there will be three independent singlet long-range modes and in that case $\alpha=3/2$. For finite albeit small interband
scattering 
%mv:
we find $\alpha\sim 1/2$ from Eq.~(\ref{alpha}). We note that the value of $\alpha$ is not universal due to asymmetry between of parameters for $\Gamma$ and $X(Y)$ pockets. This non-universality resembles non-universal weak localization correction in a metal with partially polarized magnetic impurities when the rotational symmetry is broken.~\cite{VavilovGlazman2003}
It is important, however, to keep in mind that as temperature is decreased, one may expect a
crossover behavior from $\alpha\sim 3/2$, when $\tau_\phi^{-1}\sim 4eDB_{\gamma}$ to $\alpha\sim 1/2$ for $\tau_\phi^{-1}\gg 4eDB_{\gamma}$, {see Fig.~2.} 

\begin{figure}[htb]
%mv: why is the function sharply peaked, while eq 1.3 states \delta \sigma ~B^2 or is horizontal scale multiplied by 10^3?
\includegraphics[width=8.25cm,height=6.25cm]{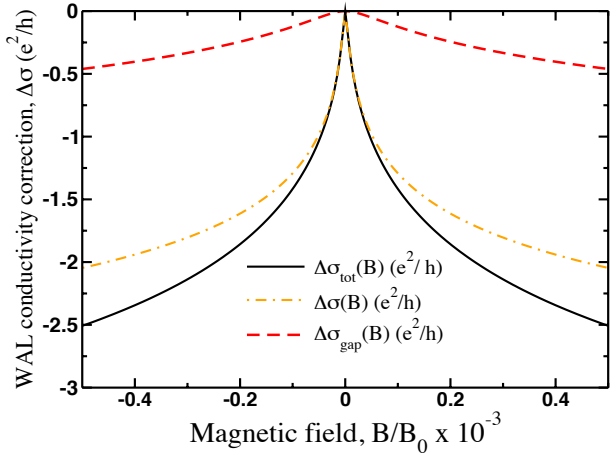}
\caption{(Color online)  Plot of the total correction to conductivity $\Delta\sigma_{\textrm{tot}})$ as a function of magnetic field for $\alpha=0.45$. The values of the scattering rates and diffusion constants have been chosen as follows: $\tau_{\Gamma Y}=1.2\tau_{\Gamma X}$, $\tau_{XY}=\tau_{\Gamma Y}$ and $D_X=D_Y=0.95D_\Gamma$.}
\label{Fig4}
\end{figure}

\subsection{WAL in correlated disorder}
It will be instructive to consider the model for the mixture of correlated disorder, when the disorder potential at a given impurity site can scatter within the same valley as well as between different valleys, and compare the corresponding value of the weight factor in this case with the one found above (\ref{alpha}).

For the correlated disorder mixture we consider the following model:
\beg\label{CorrDisorder}
V({\vec r})=\sum\limits_{i}{u}\hat{U}\delta({\vec r}-{\vec R}_i), \quad 
\hat{U}=\hat{U}_0+\zeta\hat{U}_x,
\en
where $\hat{U}_0$, $\hat{U}_x$ are defined in Eq. (\ref{Uinter}) and coefficient $\zeta=u_x/u$. Now in addition to the 
two scattering times $\tau_{a0}$ and $\tau_{a}$ defined previously (\ref{taua0},\ref{sctime}) we also introduce the relaxation time due to disorder correlations:
\beg\label{NewTimes}
\tau_{ax}^{-1}=\pi n_i\zeta u^2\nu_a=\zeta \tau_{a0}^{-1}.
\en
In turn, the self-energy matrix becomes off-diagonal in band indices due to disorder correlations:
\beg\label{SigmaNew}
\hat{\Sigma}_{ab}^{R,A}(\epsilon)\approx\mp \frac{i\hat{\sigma}_0}{2\tau_{a}}\mp\frac{i\hat{U}_x}{2\tau_{ax}}.
\en
As a consequence of this, the matrix Green's function $\check{G}^{-1}=\check{G}_0^{-1}-\check{\Sigma}$ also becomes non-diagonal in band indices. However, in the computation of the Cooperon matrix within the required accuracy we can neglect the presence of the $\tau_{ax}^{-1}$ in the Green's functions. The reason is that in the calculation of the singlet components of the Cooperon, the matrix elements which are proportional to $\tau_{ax}^{-1}$ are of the order of $(p_Fl)^{-1}\ll 1$ and therefore can be neglected. 

The analysis of the Cooperon eigenvalues can now be done along the same lines as above. Specifically, the right hand side of the equation (\ref{Eigen1}) will now acquire an extra term proportional to $\tau_{ax}^{-1}$. Setting the eigenvalue $\lambda_0=1$ and solving for the eigenvectors within the required accuracy yields:
\beg\label{neweigen}
\begin{split}
&\Psi_{ab}^{(0)}=\frac{\tilde{\tau}^*}{\tilde{\tau}_a}\delta_{ab} + (1-\delta_{ab})\left(\frac{\tau_a}{\tau_{ax}}\frac{\tilde{\tau}^*}{\tilde{\tau}_a}+\frac{\tau_b}{\tau_{bx}}\frac{\tilde{\tau}^*}{\tilde{\tau}_b}\right), 
\end{split}
\en
where $\tilde{\tau}^*\sim\tau^*$ appears as a result of normalization of the eigenvectors. An explicit expression for it in terms
of scattering times is quite cumbersome and will not be given here. In addition,  we also introduced
\beg\label{tildetaua}
\begin{split}
\tilde{\tau}_a=\tau_a\left[1-\frac{\tau_{\textrm{inter}}}{\tau_{a0}}\left(\frac{2\tau_a}{\tau_{ax}}+\sum\limits_{b\not=a}\frac{\tau_b}{\tau_{bx}}\right)\right], 
\end{split}
\en
and $\tau_{\textrm{inter}}^{-1}=\tau_{\Gamma X}^{-1}+\tau_{X Y}^{-1}+\tau_{Y\Gamma}^{-1}$ stands for the average interband scattering rate. 

From our results for the eigenvectors it is clear that the finite $\tau_{ax}$ scattering time introduces off-diagonal in valley indices components of the Cooperon matrix, which give non-zero contribution to conductivity. 
However, the contribution of these off-diagonal terms is sub-leading in power of $\tau_a/\tau_{ax}$ to the diagonal ones. If the off-diagonal Cooperon elements are neglected, the resulting expression for the coefficient $\alpha$ in the HLN formula reads
\beg\label{alphax}
\alpha_x=\sum\limits_{a}\frac{\widetilde{D}_a}{2\widetilde{D}}\left(2-\frac{\tau_{a}}{\tau_{a0}}\right)\frac{\tilde{\tau}_t}{\tau_{a0}},
\en
where $\widetilde{D}_a=v_a^2\tilde{\tau}_a$ and $\widetilde{D}$ is found by the same expression as in (\ref{LamD}) by replacing $\tau_{t,a}$ with $\tilde{\tau}_{t,a}$ and $\tilde{\tau}_t^{-1}=\tilde{\tau}_\Gamma^{-1}+\tilde{\tau}_X^{-1}+\tilde{\tau}_Y^{-1}$.
As we expected, the finite $\tau_{ax}$ further reduce the weak-antilocalization effect. 

% Discussion section %
%mv: can we join these two sections together?
\section{Discussion and Conclusions}\label{Discussion}
Perhaps one of the main challenges in identifying the nature of the conducting surface states, such as the dispersion and helicity, in topological Kondo insulator such as 
%mv: s such as 
SmB$_6$ lies in establishing to what extent the Dirac surface states remain well defined despite the fact that (a) hybridization between the conduction $d$- and $f$-orbitals is significantly reduced on the surface \cite{Onur2015} and (b) the band bending effects due to disorder scattering on the surface, which leads to an appearance of the states inside the hybridization gap. \cite{Roy2014} For example, recent ARPES  measurements (see Ref.~\onlinecite{Oliver2015} and references therein) and low-frequency, radio-frequency and microwave conductivity \cite{GlushkovInGap} seem to be in support of the picture in which conventional Rashba-split bands dominate the low-temperature transport properties on the surface of SmB$_6$. On the other hand, recent radiation spectroscopy measurements and magneto-thermoelectric transport results and well as spin-resolved ARPES \cite{Xu2014} support the picture of the topologically protected surface states. Interestingly, the Nernst effect data on the (011) plane reports the effective mass for
%mv: for 
the carries of the order of 100 of bare electron mass in agreement with existing theoretical estimates.\cite{Alexandrov2013,Roy2014}

Observation of the weak anti-localization correction to conductivity in topological insulators generally serves as an indication of the strong-spin orbit coupling and, therefore, is used to confirm the helicity of the conducting surface states. 
In topological Kondo insulators, however, the analysis of the experimental data is complicated by the possibility of the conventional polar bands which will be split by the Bychkov-Rashba spin-orbit coupling $\lambda_{SO}$. When disorder does not induce
scattering between the Dirac and parabolic bands, one may expect a weak-antilocalization correction to conductivity provided the spin-orbit coupling is strong enough. \cite{Skvortsov1998,Pikusetal1998}. Furthermore, the correction appears to be non-universal and
is proportional to $(\lambda_{SO}p_F\tau_{\textrm{tr}})^{-2}$ where $\tau_{\textrm{tr}}$ is a transport time.\cite{Skvortsov1998} Clearly, whether the magnitude of this correction is of the same order as the one we find for Dirac electrons depends on the magnitude of the $\lambda_{SO}$: at strong Bychkov-Rashba splitting of the parabolic bands their contribution to conductivity quantum correction may be strongly suppressed by this additional factor. Lastly, for the scattering which mixes the Dirac bands with the parabolic bands, one single diffusion mode is expected to be present leading to $\alpha\sim 1/2$ for a given surface. 

%mv: Another correction to our results above may appear due to the presence of .
Another correction to our results above may appear due to the presence of magnetic scattering on the surface, which may change the picture of weak localization corrections presented above. In fact, based on the magnetoresistance data \cite{JP1d}, it has been recently argued that unscreened $f$-electrons give rise to ferromagnetic state on the surface of SmB$_6$. 
%mv: One expects 
One expects therefore, that the spin-flip scattering on the surface will gap out all Dirac bands completely suppressing transport. Even in the case when only one Dirac pocket is not gapped, the quantum interference correction to conductivity will be strongly suppressed resulting in the values of $\alpha<1/2$. In addition, as it has been shown recently, the opening of the gap in the Dirac bands may actually change the sign in the quantum correction to conductivity from
weak-antilocalization to weak localization.\cite{HZLu2011} Perhaps the fact that this crossover has not been observed \cite{JP1d} suggests that at least one of the Dirac bands remains ungapped. 

%mv combine Disc and Conc  shall we use present/past time for verbs?  
In this paper, we have
considered the quantum correction to conductivity in the model with three Dirac bands with intraband and
interband disorder. We 
%find is also Ok
find that for the case of the interband disorder there is only one singlet long-range mode leading to the 
quantum correction to conductivity. The resulting expression for the weight factor shows that depending on the ratio between the
diffusion coefficients and interband scattering times, one may expect the smooth crossover from $\alpha\sim 3/2$ to $\alpha\sim 1/2$. 
In contrast with the single band case when $\alpha=1/2$, the presence of the interband scattering reduces the value of $\alpha$ so that it becomes non-universal. 
In fact, the interband scattering itself may be asymmetric, due to the large Fermi line missmatch between $\Gamma$ and $X,Y$ bands the scattering between them could be suppressed. In this case,
independent contribution of the $\Gamma$ in addition to the mixed $X$ and $Y$ components would result in $\alpha\approx1$ which may explain the value observed in Ref.~\onlinecite{Thomas2013WAL}.
Our results are generally in agreement with recent low-temperature transport experiments\cite{Thomas2013WAL,JP1d} on SmB$_6$.
%mv: added   , do we need it?   \cite{Thomas2013WAL,JP1d}
\section*{Acknowledgments}
This work was financially supported by KSU and MPI-PKS (M.D.), DARPA and Simons Foundation (V.G.), NSF Grant No. DMR-0955500 (M.V.) We thank Igor Aleiner, Leonid Glazman, Alex Levchenko and Jing Xia for useful discussions.

\begin{widetext}
\appendix
\section{Components of the matrix $\hat{M}$}\label{MIntra}
In this Section we provide an explicit derivation for the elements of the matrix $\hat{M}$ which enters
into the equation for the Cooperon (\ref{Eq4C}).
This matrix is defined as follows
\beg
\begin{split}
&M_{S_1S}^{ab}(\omega,\bq)=\frac{1}{2}\int\frac{d^2\bp}{4\pi^2}\textrm{Tr}\left\{[\hat{G}_a^R(\epsilon+\omega,\bp)]^T(\sigma_y\sigma_{S_1})
[\hat{G}_b^A(\epsilon,\bq-\bp)](\sigma_{S}\sigma_{y})\right\}
\end{split}
\en
First, we compute the diagonal components in pseudospin space:
\beg\label{Mab}
\begin{split}
M_{00}^{ab}(\omega,\bq)&=\int\frac{d^2\bp}{4\pi^2}\left\{\frac{(\epsilon+\omega+i/2\tau_a)(\epsilon-i/2\tau_b)+v_av_bp^2-v_av_b\bp\bq}{[(\epsilon+\omega+i/2\tau_a)^2-v_a^2p^2][(\epsilon-i/2\tau_b)^2-v_b^2(\bq-\bp)^2]}\right\}, \\
M_{11}^{ab}(\omega,\bq)&=\int\frac{d^2\bp}{4\pi^2}\left\{\frac{(\epsilon+\omega+i/2\tau_a)(\epsilon-i/2\tau_b)+v_av_b[p_y(q_y-p_y)-p_x(q_x-p_x)]}{[(\epsilon+\omega+i/2\tau_a)^2-v_a^2p^2][(\epsilon-i/2\tau_b)^2-v_b^2(\bq-\bp)^2]}\right\}, \\
M_{22}^{ab}(\omega,\bq)&=\int\frac{d^2\bp}{4\pi^2}\left\{\frac{(\epsilon+\omega+i/2\tau_a)(\epsilon-i/2\tau_b)+v_av_b[p_x(q_x-p_x)-p_y(q_y-p_y)]}{[(\epsilon+\omega+i/2\tau_a)^2-v_a^2p^2][(\epsilon-i/2\tau_b)^2-v_b^2(\bq-\bp)^2]}\right\}, \\
M_{33}^{ab}(\omega,\bq)&=\int\frac{d^2\bp}{4\pi^2}\left\{\frac{(\epsilon+\omega+i/2\tau_a)(\epsilon-i/2\tau_b)-v_av_bp^2+v_av_b\bp\bq}{[(\epsilon+\omega+i/2\tau_a)^2-v_a^2p^2][(\epsilon-i/2\tau_b)^2-v_b^2(\bq-\bp)^2]}\right\}.
\end{split}
\en
We calculate each of these integrals separately. In what follows I use the following approximation
\beg
\omega\tau_{a,b}\ll 1, \quad v_j|\bq-\bp|\approx v_jp\left(1-\frac{q\cos\phi}{p_{Fj}}+\frac{q^2\sin^2\phi}{2p_{Fj}^2}\right)
\en
Consider the following integral
\beg\label{Generic}
\begin{split}
I_{ab}=\int\frac{d^2\bp}{4\pi^2}\left\{\frac{1}{[(\epsilon+\omega+i/2\tau_a)^2-v_a^2p^2][(\epsilon-i/2\tau_b)^2-v_b^2p^2]}\right\}
\end{split}
\en
and introduce the following variables $\xi=\textrm{sign}(\epsilon)v_ap-\epsilon$, $r_{ab}=v_a/v_b$. We have
\beg
\begin{split}
&I_{ab}=\frac{1}{2\pi v_b^2}\int\limits_{-\epsilon}^{\textrm{sign}(\epsilon)\infty}\frac{[\textrm{sign}(\epsilon)\xi+|\epsilon|]\textrm{sign}(\epsilon)d\xi}{[(\epsilon+\omega+i/2\tau_a)^2-v_a^2p^2][r_{ab}^2(\epsilon-i/2\tau_b)^2-v_a^2p^2]}\approx\\&\approx
\frac{|\epsilon|}{2\pi v_b^2}\frac{1}{2\epsilon^2(1+r_{ab})}\int\limits_{-\infty}^{+\infty}\frac{d\xi}{(\omega+i/2\tau_a-\xi)[(r_{ab}-1)\epsilon-i/2\tau_b-\xi]}\approx\\&\approx\frac{2\pi\nu_b\tau_{ab}}{1-2i\omega\tau_{ab}-2i(1-r_{ab})\epsilon\tau_{ab}}, \quad 
\tau_{ab}^{-1}=\tau_a^{-1}+\tau_b^{-1}.
\end{split}
\en
Clearly, from this expression it follows that all off-diagonal elements of the Cooperon matrix will 
remain finite in zero momentum and frequency limit. Thus we need to analyze (\ref{Mab}) for $a=b$ only. 
We have
\beg
\begin{split}
&M_{00}^{aa}\approx 1+i\omega\tau_a-\frac{v_{a}^2q^2\tau_a^2}{2}, \quad M_{33}^{aa}\approx -\frac{\omega+\frac{i}{\tau_a}}{2v_ap_{Fa}}\ll 1, \\
&M_{11}^{aa}\approx\frac{1}{2}\left[1+i\omega\tau_a-\frac{v_{a}^2q^2\tau_a^2}{2}-\frac{v_{a}^2q^2\tau_a^2\cos(2\varphi_q)}{4}
\right], \quad M_{22}^{aa}\approx\frac{1}{2}\left[1+i\omega\tau_a-\frac{v_{a}^2q^2\tau_a^2}{2}+\frac{v_{a}^2q^2\tau_a^2\cos(2\varphi_q)}{4}
\right],
\end{split}
\en
where we used 
\beg
\pi\nu_a\tau_a\Lambda=1. 
\en
Next, we consider the off-diagonal components is pseudospin space. We have:
\beg
\begin{split}
M_{01}^{ab}(\omega,\bq)&=\int\frac{d^2\bp}{4\pi^2}\left\{\frac{(\epsilon+\omega+i/2\tau_a)(q_x-p_x)+(\epsilon-i/2\tau_b)p_x}{[(\epsilon+\omega+i/2\tau_a)^2-v_a^2p^2][(\epsilon-i/2\tau_b)^2-v_b^2(\bq-\bp)^2]}\right\}, \\
M_{02}^{ab}(\omega,\bq)&=\int\frac{d^2\bp}{4\pi^2}\left\{\frac{(\epsilon+\omega+i/2\tau_a)(q_y-p_y)+(\epsilon-i/2\tau_b)p_y}{[(\epsilon+\omega+i/2\tau_a)^2-v_a^2p^2][(\epsilon-i/2\tau_b)^2-v_b^2(\bq-\bp)^2]}\right\}, \\
M_{12}^{ab}(\omega,\bq)&=-\int\frac{d^2\bp}{4\pi^2}\left\{\frac{(\epsilon+\omega+i/2\tau_a)p_y(q_x-p_x)+(\epsilon-i/2\tau_b)(q_y-p_y)p_x}{[(\epsilon+\omega+i/2\tau_a)^2-v_a^2p^2][(\epsilon-i/2\tau_b)^2-v_b^2(\bq-\bp)^2]}\right\}, \\
\end{split}
\en
while the remaining components will give zero. Thus, collecting all the terms we obtain:
\beg\label{M}
\hat{M}=\left[
\begin{matrix}
\frac{v_{a}^2q^2\tau_a^2}{2}-i\omega\tau_a & -\frac{i}{2}v_aq\tau_a\cos\varphi_q &-\frac{i}{2}v_aq\tau_a\sin\varphi_q & 0 \\
-\frac{i}{2}v_aq\tau_a\cos\varphi_q & \frac{1}{2}\left(1-i\omega\tau_a+\frac{v_{a}^2q^2\tau_a^2}{2}+\frac{v_{a}^2q^2\tau_a^2\cos(2\varphi_q)}{4}\right) & \frac{1}{8}v_a^2q^2\tau_a^2\sin(2\varphi_q) & 0 \\
-\frac{i}{2}v_aq\tau_a\sin\varphi_q & -\frac{1}{2}v_a^2q^2\tau_a^2\sin(2\varphi_q) & \frac{1}{2}\left(1-i\omega\tau_a+\frac{v_{a}^2q^2\tau_a^2}{2}-\frac{v_{a}^2q^2\tau_a^2\cos(2\varphi_q)}{4}\right) &  0 \\
 0 & 0 & 0 & 1
\end{matrix}
\right].
\en
This result shows that only single mode corresponding to the singlet component of the Cooperon matrix will remain gapless. 
\section{Quantum corrections to conductivity}\label{ConductivityCorrections}
\subsection{bare Hikami box}
\begin{figure}[htb]
%mv: why is the function sharply peaked, while eq 1.3 states \delta \sigma ~B^2 or is horizontal scale multiplied by 10^3?
\includegraphics[width=11.25cm,height=4.25cm]{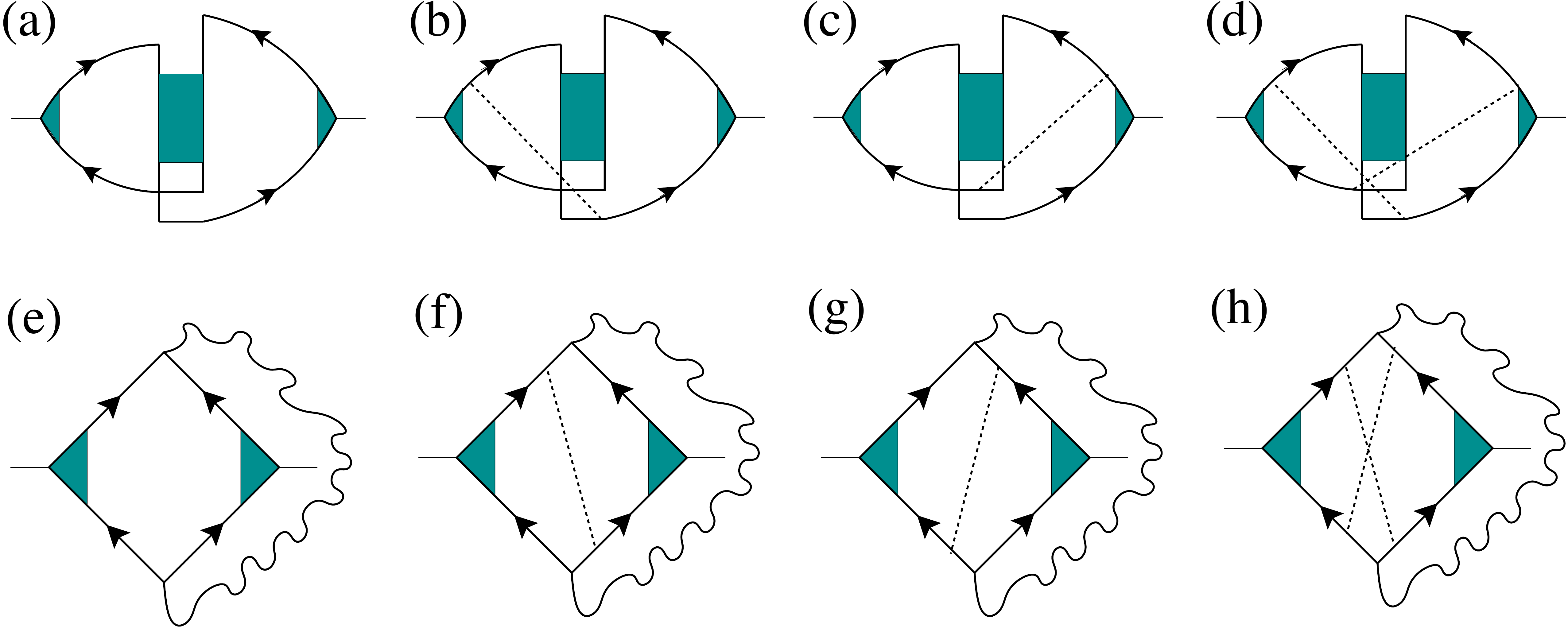}
\caption{\label{Fig7}(Color online)  Panels (a)-(d): diagrams contributing to the weak localization correction to conductivity. Panels (e)-(h): the same contributions as (a)-(d), shown in the representation of Hikami boxes.}
\end{figure}

Contribution to conductivity from the bare Hikami box, Fig. \ref{Fig7}(a,e),
 is:
\beg\label{sigmaHikami}
\begin{split}
\delta\sigma_{ij}^{(1)}=\frac{e^2}{2\pi}\int\frac{d\bk d\bq}{(2\pi)^4}\sum\limits v_av_b[\hat{G}_a^A(\bk,\epsilon)\sigma_i\hat{G}_a^R(\bk,\epsilon+\omega)]_{\alpha\beta}\Gamma_{\beta\delta\gamma\alpha}^{ba,ba}(\omega,\bq)[\hat{G}_b^R(\bq-\bk,\epsilon+\omega)\sigma_j\hat{G}_b^A(\bq-\bk,\epsilon)]_{\gamma\delta}
\end{split}
\en
To evaluate this correction we employ Eq. (\ref{GammaToCs1s2}). Calculation of the trace over the pseudospin degrees of freedom is done by Mathematica. The resulting expression can be simplified by neglecting the dependence on external momentum $\bq$ in the single particle
correlators. In addition, as it follows from the calculation of the traces we can also neglect the frequency dependence in the denominators. 
We are interested in find the contribution from the most singular terms in the Cooperon. For the diagonal components of conductivity it follows
\beg\label{Sigma1}
\begin{split}
\delta\sigma_{ii}^{(1)}=\sum\limits_{a=\Gamma,X,Y}\frac{e^2v_a^2}{2\pi}\int\frac{d\bk d\bq}{(2\pi)^4}\sum\limits_{S_1S_2}
\frac{1}{2}C_{S_1S_2}^a(\bq)&\textrm{Tr}\left\{\hat{G}_a^A(\bk,\epsilon)\sigma_i\hat{G}_a^R(\bk,\epsilon+\omega)\sigma_y^T\sigma_{S_1}^T\times\right.\\ &\times\left.[\hat{G}_a^A(\bq-\bk,\epsilon)]^T\sigma_i^T[\hat{G}_a^R(\bq-\bk,\epsilon+\omega)]^T\sigma_y\sigma_{S_2}\right\}
\end{split}
\en
To calculate the trace we use the following relations:
\beg\label{Kostya}
[\hat{G}_a^{R,A}(\bk,\epsilon)]^T=\frac{\sigma_y[\epsilon_{R,A}\sigma_0-v_a(k_x\sigma_x+k_y\sigma_y)]\sigma_y}{\epsilon_{R,A}^2-v_a^2k^2}, \quad \sigma_{i}^T=[2\delta_{i,0}-1]\sigma_y\sigma_i\sigma_y, \quad (i=0,x,y,z).
\en
Using these relations and neglecting the $q$ dependence in the nominators of the $G^{A,R}$ for the trace we find 
\beg
\begin{split}
\frac{1}{2}\int\limits_{0}^{2\pi}\frac{d\phi}{2\pi}&\textrm{Tr}\left\{\hat{G}_a^A(\bk,\epsilon)\sigma_i\hat{G}_a^R(\bk,\epsilon+\omega)\sigma_y^T\sigma_{S_1}^T[\hat{G}_a^A(\bq-\bk,\epsilon)]^T\sigma_i^T[\hat{G}_a^R(\bq-\bk,\epsilon+\omega)]^T\sigma_y\sigma_{S2}\right\}\approx\\ &\approx
\delta_{S_1S_2}\epsilon^4\left\{4\delta_{S_1,0}-3\delta_{S_1,i}-\delta_{S_1,\overline{i}}\right\}
\end{split}
\en
where on the last step I used $\omega\ll\epsilon$ and I have also assumed $v_ak\approx\epsilon$. Thus (\ref{Sigma1}) becomes
\beg
\begin{split}
\delta\sigma_{ii}^{(1)}=\sum\limits_{a=\Gamma,X,Y}\frac{e^2v_a^2}{2\pi}\int\frac{d\bq}{(2\pi)^2}\left[4C_{00}^a(\bq)-3C_{ii}^a(\bq)-C_{\overline{i}\overline{i}}^a(\bq)\right]\int\frac{kdk}{2\pi}\frac{\epsilon^4}{[(\epsilon_R+\omega)^2-v_a^2k^2]^2[\epsilon_A^2-v_a^2k^2]^2}
\end{split}
\en
We deal with the momentum integral as follows:
\beg
\begin{split}
\int\limits_0^\infty\frac{kdk}{2\pi}f(vk)=&\frac{1}{2\pi v^2}\int\limits_{-\epsilon}^{\textrm{sign}(\epsilon)\infty}\left\{\textrm{sign}(\epsilon)\varepsilon_k+|\epsilon|\right\}
\textrm{sign}(\epsilon)d\varepsilon_kf\left((\epsilon+\varepsilon_k)\textrm{sign}(\epsilon)\right)\approx\\&\approx
\frac{|\epsilon|}{2\pi v^2}\int\limits_{-\infty}^{+\infty} f\left(\xi+|\epsilon|\right)d\xi=\nu\int\limits_{-\infty}^{+\infty} f\left(\xi+|\epsilon|\right)d\xi
\end{split}
\en
Finally, the result is
\beg\label{first}
\begin{split}
\delta\sigma_{ii}^{(1)}=\sum\limits_{a=\Gamma,X,Y}\frac{e^2\nu_av_a^2\tau_a^3}{8(1-i\omega\tau)^3}\int\frac{d\bq}{(2\pi)^2}\left[4C_{00}^a(\bq)-3C_{ii}^a(\bq)-C_{\overline{i}\overline{i}}^a(\bq)\right].
\end{split}
\en
\subsection{first disorder correction to the Hikami box}
The expression for the second correction to conductivity, shown on Fig. \ref{Fig7}(b,f), reads
\beg\label{Sigma2}
\begin{split}
\delta\sigma_{ii}^{(2)}=&\Lambda_0\sum\limits_{a=\Gamma,X,Y}\frac{e^2v_a^2}{2\pi}\int\frac{d\bk d\bp d\bq}{(2\pi)^6}\sum\limits_{S_1S_2}
\frac{1}{2}C_{S_1S_2}^a(\bq)\textrm{Tr}\left\{\hat{G}_a^A(\bk,\epsilon)\sigma_i\hat{G}_a^R(\bk,\epsilon+\omega)
\hat{U}\hat{G}_a^R(\bp,\epsilon+\omega)
\sigma_y^T\sigma_{S_1}^T\times\right.\\ &\times\left.[\hat{G}_a^A(\bq-\bp,\epsilon)]^T\sigma_i^T[\hat{G}_a^R(\bq-\bp,\epsilon+\omega)]^T\hat{U}^T[\hat{G}_a^R(\bq-\bk,\epsilon+\omega)]^T\sigma_y\sigma_{S_2}\right\}.
\end{split}
\en
Here we took into account that only diagonal part of the disorder potential contributes to the conductivity correction, since correlation functions are diagonal in valley indices. Taking into account the expressions for the propagators and relations (\ref{Kostya}) we obtain
\beg\label{Sigma2p}
\begin{split}
\delta\sigma_{ii}^{(2)}=&\Lambda_0\sum\limits_{a=\Gamma,X,Y}\frac{e^2v_a^2}{2\pi}\int\frac{d\bq}{4\pi^2}\sum\limits_{S_1S_2}
\left(\delta_{S_1,0}-\frac{1}{2}\right)C_{S_1S_2}^a(\bq)\times\\&\int\frac{pdp}{2\pi}\frac{1}{[(\epsilon_R+\omega+\frac{i}{2\tau})^2-v_a^2p^2]^2[(\epsilon-\frac{i}{2\tau})^2-v_a^2p^2]}\int\frac{kdk}{2\pi}\frac{1}{[(\epsilon_R+\omega+\frac{i}{2\tau})^2-v_a^2k^2]^2[(\epsilon-\frac{i}{2\tau})^2-v_a^2k^2]}\\&\times
\textrm{Tr}\left\{[\epsilon_A\sigma_0+v_a(\bk\cdot{\vec \sigma})]\sigma_i[(\epsilon_R+\omega)\sigma_0+v_a(\bk\cdot{\vec \sigma})]
[(\epsilon_R+\omega)\sigma_0+v_a(\bp\cdot{\vec \sigma})]
\sigma_{S_1}\times\right.\\ &\times\left.[\epsilon_A\sigma_0+v_a(\bp\cdot{\vec \sigma})]\sigma_i[(\epsilon_R+\omega)\sigma_0+v_a(\bp\cdot{\vec \sigma})][(\epsilon_R+\omega)\sigma_0+v_a(\bk\cdot{\vec \sigma})]\sigma_{S_2}\right\}.
\end{split}
\en
Computation of the trace and subsequent integration over momenta in the limit $\omega=0$ yields
\beg\label{second}
\delta\sigma_{ii}^{(2)}=-\sum\limits_{a=\Gamma,X,Y}\frac{e^2\nu_av_a^2\tau_a^3}{8}\left(\frac{\tau_{a}}{\tau_{a0}}\right)\int\frac{d\bq}{(2\pi)^2}\left[C_{00}^a(\bq)-C_{ii}^a(\bq)\right].
\en
Note, that an additional pre-factor appears since the relation time differs from the intra-pocket scattering time 

\subsection{second disorder correction to the Hikami box}
The third correction, Fig. \ref{Fig7}(c,g), is the same as the first correction (\ref{second}) to the Hikami box diagram:
\beg\label{third}
\begin{split}
\delta\sigma_{ii}^{(3)}&=-\sum\limits_{a=\Gamma,X,Y}\frac{e^2\nu_av_a^2\tau_a^3}{8}\left(\frac{\tau_{a}}{\tau_{a0}}\right)\int\frac{d\bq}{(2\pi)^2}\left[C_{00}^a(\bq)-C_{ii}^a(\bq)\right].
\end{split}
\en
Finally, the fourth and the last correction to conductivity, Fig. \ref{Fig7} (d,g), is small in parameter $1/p_Fl\ll 1$ and can be ignored. 
Adding up all three contributions to the conductivity we find expression (\ref{TotalCorrection}) in the main text.
\end{widetext}

\bibliography{smb6wal}

\end{document}